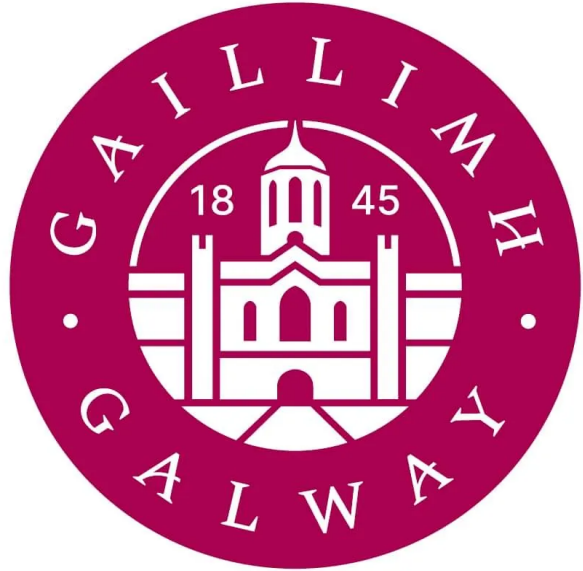

# Wavelet-Based Time-Frequency Fingerprinting for Feature Extraction of Traditional Irish Music

Noah Shore

# OLLSCOIL NA GAILLIMHE
# UNIVERSITY OF GALWAY

# Wavelet-Based Time-Frequency Fingerprinting for Feature Extraction of Traditional Irish Music

## Noah Shore

Thesis submitted in fulfilment of the requirements for the degree of
**Master of Science (MSc)**
**Supervisor:** Dr. Michael Mc Gettrick
**Head of School:** Prof. Cathal Seoighe

Ollscoil na Gaillimhe
―――――――――――――
University of Galway
School of Mathematical and Statistical Sciences

July, 2025

*This work presents a wavelet-based approach to time-frequency fingerprinting for time series feature extraction, with a focus on audio identification from live recordings of traditional Irish tunes. The challenges of identifying features in time-series data are addressed by employing a continuous wavelet transform to extract spectral features and wavelet coherence analysis is used to compare recorded audio spectrograms to synthetically generated tunes. The synthetic tunes are derived from ABC notation, which is a common symbolic representation for Irish music. Experimental results demonstrate that the wavelet-based method can accurately and efficiently identify recorded tunes. This research study also details the performance of the wavelet coherence model, highlighting its strengths over other methods of time-frequency decomposition. Additionally, we discuss and deploy the model on several applications beyond music, including in EEG signal analysis and financial time series forecasting.*

# Contents

# Abbreviations

| | |
|---|---|
| **ABC** | ABC Notation |
| **AIRI** | All-India Rainfall Index |
| **BPM** | Beats per Minute |
| **CL** | Colgate-Palmolive Co |
| **COI** | Cone of Influence |
| **CWT** | Continuous Wavelet Transform |
| **dB** | Decibels |
| **DWT** | Discrete Wavelet Transform |
| **EEG** | Electroencephalogram |
| **ES** | Eversource Energy |
| **fCWT** | Fast Continuous Wavelet Transform |
| **FT** | Fourier Transform |
| **FFT** | Fast Fourier Transform |
| **Fz** | Frontal Midline Electrode |
| **Hz** | Hertz |
| **iFFT** | Inverse Fast Fourier Transform |
| **kHz** | Kilohertz |
| **MATT2** | Machine Annotation of Traditional Tunes |
| **Niño 3** | El Niño-Southern Oscillation Region 3 |
| **NYSE** | New York Stock Exchange |
| **Pz** | Parietal Midline Electrode |
| **S** | Seconds |
| **SR** | Sampling Rate |
| **SST** | Sea Surface Temperature |
| **STFT** | Short-Time Fourier Transform |
| **TFR** | Time-Frequency Representation |
| **WCA** | Wavelet Coherence Analysis |
| **XWT** | Cross-Wavelet Transform |

6 Contents

# Chapter 1

# Introduction

> The opus, along with its principal examples, is outlined.

## 1.1 Overview

The information curated for this opus is organized into four main chapters, an introduction, and two appendices. In the opening chapter, an overview is laid out of the objectives and methodology in this research, as well as an apologia for the use of traditional Irish music as a testing ground for the aforementioned methodology. These upcoming sections aim to lay the groundwork for the following chapter.

Chapter 2 delves into the mathematical prerequisites to be utilized later in the work, providing preliminary context for some of the back-end processes leading up to grander modules. This chapter opens with a discussion of Heisenberg's uncertainty principle, which implicates some difficult compromises in time-frequency analysis and serves as motivation for the dynamic decompositions to be explored. Following this is waveforms, frequencies, and Fourier transforms, which will lead into the next section on sound waves and frequency analysis. We then look at time-dependent decompositions of sound, beginning with short-time Fourier transforms and Gabor transforms, before introducing the favored method in this application: the wavelet transform.

In Chapter 3, we look at the methods used to produce the musical examples and obtain results. This chapter begins with an introduction of the data formats, including live audio as well as symbolic sheet music. The next section shows the data pipeline for converting the symbolic music into a waveform that can be compared with the recording. We then use these datasets to produce spectrograms via the continuous wavelet transform, and finishing up by introducing the wavelet coherence formula and displaying the coherence between the symbolic and recorded time series.

We use the coherence in Chapter 4 to construct the tune identification model and obtain results. This chapter opens with a discussion of the comparison between spectrograms and displays the results for a recording compared with a database of tunes. This is followed up with a section on factors that impact the prediction accuracy of the model, showing the improved results after some of these modifications are implemented. We then have a discussion of the model's reaction to instrumentation, showing how the results differ for the flute vs the fiddle. The chapter closes with an investigation into the model's biases, where two control data points are fabricated to examine how the model favors some tunes, and what it is likely to choose when the input data is meaningless.

As a conclusion, Chapter 5 serves to reopen some topics alluded to by the methodology in Chapters 2 and 3 but unrelated to the tune identifier model. This includes a discussion of the phase products of the coherence model, followed by applications of these phases in electroencephalography and econometrics. The concluding section of this chapter contains some closing remarks and possibilities for future work on the topics discussed.

Appendix A presents some of the code used to produce the waveforms for the tune wavelet model. The first section showcases several waveform generator functions in Python, starting with the simple sine wave. Subsequently we have the piano note and banjo note generator functions as well.

Any other allusions in the text are referenced to Appendix B. This appendix opens with a look at the cone of influence, a vital component of proper wavelet transform interpretation. This section contains an example with El Niño being compared to average rainfall in India. The last section of Appendix B re-displays some of the results from Chapter 4 without a previously applied calibration to show how bias-resistant the tune-id model is.

## 1.2 Traditional Irish Music as a Case Study

Traditional Irish tunes provide an ideal testing ground to showcase the accuracy and efficiency of the methods outlined in this work. The structure of Irish reels, jigs, and other dance tunes is highly regular, making them well-suited for apples-to-apples comparisons.

For this research, only reels are used though all the same techniques would work for jigs, hornpipes, polkas, etc. with minimal changes to the settings. Reels have a time signature of 4/4, and fall into two main buckets: single reel and double reel. For the single reel, we have two parts of 8 measures each called A and B parts, which alternate back and forth. While also containing A and B parts, the double reel on the other hand has an AABB format, making each cycle twice as long. In this application, all[1] double reels are converted to singles to maintain equal length across tunes.

The symbolic representations of tunes through ABC notation are easily accessible, allowing for the generation of synthetic versions of these tunes, which can serve as a basis for comparison in the tune recognition tasks. By feeding this notation through the pipeline laid out in Chapter 3, we obtain synthesized versions of the tunes that we can run the wavelet analysis on.

All of our tune data comes from *thesession.org*, which is the default tunebase for Irish music players. Each tune has lots of metadata, including aliases, dance format, composer, and key. Dealing with this metadata is necessary for the applications in this work, and the SQL queries and dataframe constructions in Python for this application can be found in Appendix A.

---

[1] Some tunes are neither single nor double, such as *The Sailor's Bonnet,* which is a one-point-five reel with the format ABB.

## 1.3 Definitions and Terms

It is necessary to import some of the corpus-specific language used in this work.

- *Signal* refers to a function that carries information via embedded frequencies. *Time series* is a data format in which, for each point in time, there is an associated value. *Time series* is not related to the analysis definition of a *series* as a sequence of summations. For this paper, time series are treated as discrete approximations of continuous signal functions, and the words will be used interchangeably.

- In some existing literature on wavelet coherence analysis, including work by Ieracitano et al. [16], the wavelet coherence arrays are referred to as *coherograms*, indicating their analogy to spectrograms. That terminology is not used in this work, but it refers to the same object.

- The *continuous* wavelet transforms being shown in this work are clearly not calculated as continuous analytical solutions for Equation (2.5.1), rather they are numerical approximations of the integral. The term *continuous* here is meant to distinguish CWT from the *discrete wavelet transform* (DWT), whose applications are outside the scope of this work.

- The terms of scale, wavelength, and frequency refer to different mathematical objects.

    - *Wavelength* is the measurable length of a cycle.
    - *Frequency* is the number of these cycles per unit time.
    - *Scale* refers to the coefficient by which a wavelet at a central frequency is stretched or compressed.

    Although these necessarily represent different values, in the context of the wavelet transform they are often used interchangeably. For example, we might say we are looking for the wavelengths in a signal, or the frequencies in a signal, and we are speaking of the same thing. Similarly, we might say that we construct a wavelet transform with a particular resolution in scales, frequencies, or wavelengths, and be referring to the same thing.

- When discussing Irish music—or folk and dance music in general—it is often helpful to think of repeated segments as A, B, C, etc., parts. Much like rhyme schemes in poetry, this makes it easy to classify tune structures using formats such as AABB, AABBCC, or simply AB. This should not be confused with ABC notation, which is a formalized system for encoding musical scores.

# Chapter 2

# Background

The mathematical prerequisites for time-series identification and fingerprinting are detailed.

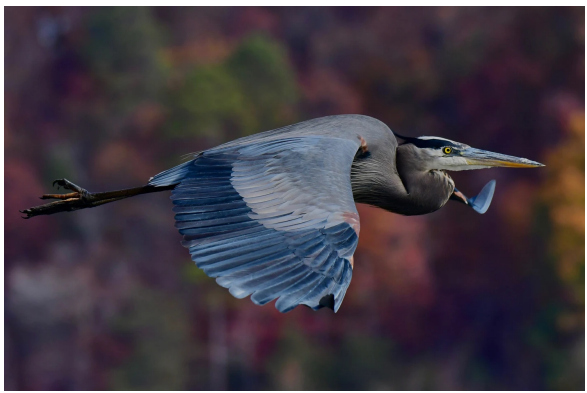

The Great Blue Heron beats its wings about 2.5 times per second [19] — a slow, steady, and controlled rhythm. This high *spectral precision* and stable momentum allow for fast and efficient flight, while sacrificing agility and maneuverability.

Christopher Grau
Audubon Photography Awards

The Ruby-Throated Hummingbird, on the other hand, ranges from 50 to 100 wingbeats per second [20], making it highly accurate with respect to time. This *time precision* allows the hummingbird to hover perfectly in place and grants it excellent aerodynamic control, though its top speed is limited and it burns through energy at unsustainable rates for long-distance flight.

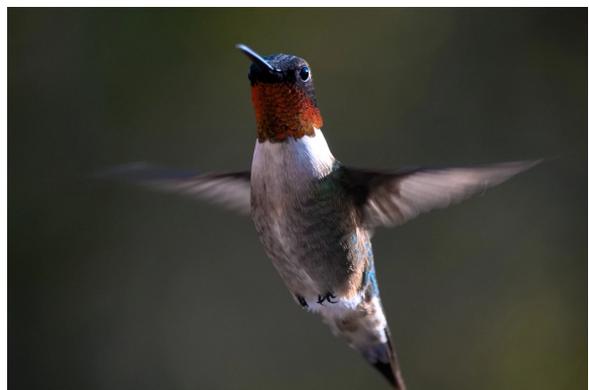

Gary Robinette
Audubon Photography Awards

## 2.1 Motivation for Dynamic Decomposition of Time Series Data

The exercise of decomposing a signal into its component frequencies has been explored since the invention of the *Fourier transform* (FT), which is fundamentally a *global* operator. This makes it perfectly capable of telling us the frequencies present without giving us any information on when they manifest. If we are interested in extracting the temporal indices of our frequencies via Fourier, we must compute transforms at multiple timesteps. Achieving accuracy in this time-frequency mesh grid quickly leads to compromising limitations derived from the uncertainty principle.

To show this, let us begin with Heisenberg's uncertainty principle [1] as it is written in the context of quantum physics:

$$\Delta x \, \Delta p \geq \frac{\hbar}{2} \tag{2.1.1}$$

For a particle, $\Delta x$ represents uncertainty in position, $\Delta p$ uncertainty in momentum, and $\hbar$ is the reduced Planck's constant.

In quantum mechanics, a particle's state can be represented as a wavefunction in either position space, $\psi(x)$, or momentum space, $\phi(p)$. These two functions are *Fourier duals*,[2] meaning the Fourier transform of $\psi(x)$ yields $\phi(p)$, and vice versa, without altering the underlying physical state. To show this duality, let us begin by producing $\phi(p)$ by the FT of $\psi(x)$.

$$\phi(p) = \frac{1}{\sqrt{2\pi\hbar}} \int_{-\infty}^{\infty} \psi(x) \, e^{-ipx/\hbar} \, dx \tag{2.1.2}$$

Conversely, the position-space wavefunction $\psi(x)$ can be recovered by the inverse Fourier transform:

$$\psi(x) = \frac{1}{\sqrt{2\pi\hbar}} \int_{-\infty}^{\infty} \phi(p) \, e^{ipx/\hbar} \, dp \tag{2.1.3}$$

In this case, when we refer to the position wavefunction vector's corresponding vector in momentum space, we mean the same quantum state, represented in a different *orthonormal basis* of the Hilbert space [3]. The Hilbert space structure provides that states can be projected onto different bases through transformations like Fourier. Likewise, in classical wave mechanics, a vector in the frequency domain is the Fourier dual of a corresponding vector in the time domain, both being representations of the same wave.

Given a wave represented by a function $f(x)$, its Fourier transform $\hat{f}(k)$ is defined as:

$$\hat{f}(k) = \frac{1}{\sqrt{2\pi}} \int_{-\infty}^{\infty} f(x) \, e^{-ikx} \, dx \tag{2.1.4}$$

The inverse Fourier transform recovers $f(x)$ from $\hat{f}(k)$:

$$f(x) = \frac{1}{\sqrt{2\pi}} \int_{-\infty}^{\infty} \hat{f}(k) \, e^{ikx} \, dk \tag{2.1.5}$$

Where $k$ is the wavenumber, proportional to frequency.

Because these are the same transform between dual spaces, the uncertainty principle is equivalently applied to time and frequency domains. This time we write it as:

$$\sigma_t \cdot \sigma_f \geq \frac{1}{4\pi} \tag{2.1.6}$$

Where $\sigma_t$ is deviation in time and $\sigma_f$ is deviation in frequency. From this formulation, we can see that more certainty in one domain comes at the cost of certainty in the other.

Returning to the implications for time-frequency decompositions, we find that fine resolutions are not affordable in both domains for many applications. If we were to create a time-frequency mesh grid over our time series and take Fourier transforms at each time step, we would have to sacrifice frequency range in order to achieve sufficient constitution in our time mesh. This type of static decomposition is known as a *Short-Time Fourier Transform* (STFT), which is defined as:

$$\text{STFT}_x(t, \omega) = \int_{-\infty}^{\infty} x(\tau)\, g(\tau - t)\, e^{-i\omega\tau}\, d\tau \tag{2.1.7}$$

Here, $x(\tau)$ is the signal, $\omega$ represents frequency, and $g$ is a window function that is convolved across the signal to reveal the frequencies present at each timestep. The STFT domain compromise is illustrated in Figure 2.1.

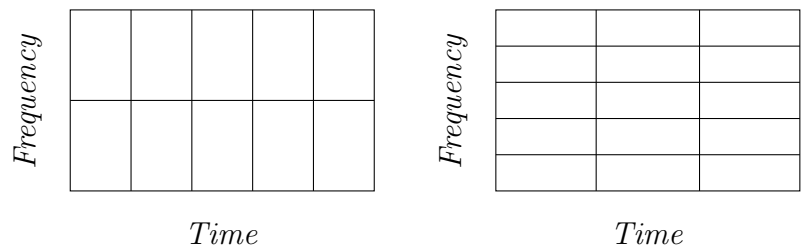

Figure 2.1: STFT tradeoff. Left: Fine time, poor frequency tiling. Right: Fine frequency, poor time tiling.

This natural compromise has led to innovation in tiling methods apart from the STFT for sensitive applications, and it has proven to be a lucrative science, as we will see in the following chapters.

## 2.2 Waveforms and the Fourier Transform

To illustrate how we pull frequency information from a waveform, let us examine a signal that was generated as a sum of multiple sine waves, as observed in Figure 2.2. For the example, we will call this signal **x**. Units of amplitude are omitted here as they are irrelevant to the example, but in time we might imagine this signal lasting for exactly 1 second.

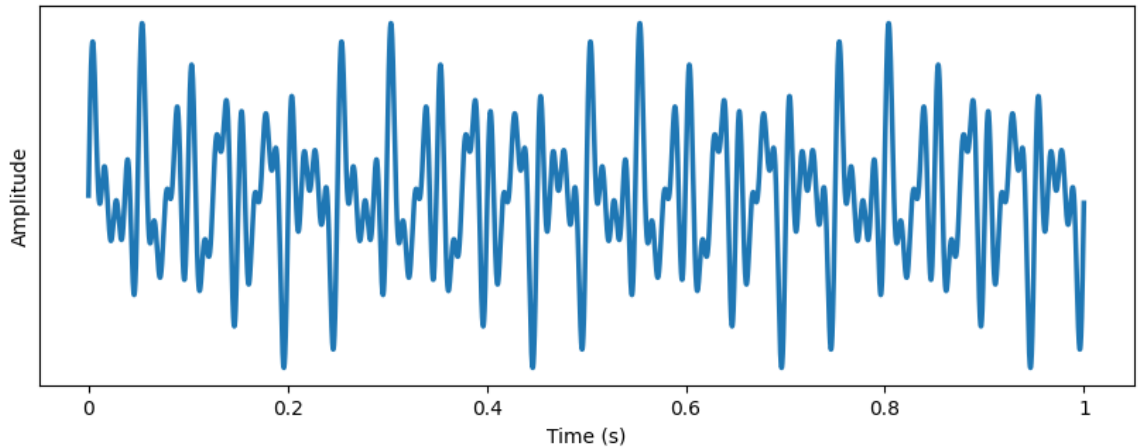

Figure 2.2: Example waveform **x**.

As we are dealing with digital signal data, **x** is not stored as a set of wavelength coefficients, but rather a time series. This means that **x** was generated with a *sampling rate*, or a chosen time-mesh grid, along which each point was sampled for amplitude.

If $d$ is the duration, $r$ is the sampling rate, and $N = r \cdot d$ is the total number of samples, then **x** is an element of $\mathbb{R}^N$. Components of **x** are properly defined as:

$$\mathbf{x} := \{x[n]\}_{n=0}^{N-1}, \quad x[n] \in \mathbb{R} \subset \mathbb{C} \tag{2.2.1}$$

Or as a column vector:

$$\mathbf{x} = \begin{bmatrix} x[0] \\ x[1] \\ x[2] \\ \vdots \\ x[N-1] \end{bmatrix} \tag{2.2.2}$$

As a vector, $\mathbf{x} \in \mathbb{R}^N \subset \mathbb{C}^N$. The $\mathbb{C}^N$ is notable here because it is the *pre-image* for the *Discrete Fourier Transform* (DFT), which is capable of processing complex data, though at this point we are dealing only with real-valued time series. Strictly speaking, DFT is a function that receives the discretized **x** as input and returns a sequence of complex coefficients:

$$DFT : \mathbb{C}^N \to \mathbb{C}^N \tag{2.2.3}$$

Whose image is defined as:

$$\mathbf{X} := \{X[k]\}_{k=0}^{N-1}, \quad X[k] \in \mathbb{C} \tag{2.2.4}$$

Where

$$X[k] = \sum_{n=0}^{N-1} x[n] \cdot e^{-i2\pi \frac{kn}{N}} \quad \text{for} \quad k = 0, 1, \ldots, N-1 \qquad (2.2.5)$$

For this example, the output **X** can be seen in Figure 2.3.

As $\mathbf{X} \in \mathbb{C}^N$ is a complex-valued vector, we are clearly missing some information when looking at the DFT output, namely, the complex arguments. When we say that Figure 2.2 (**x**) and Figure 2.3 (**X**) are different representations of the same wave, note that only the magnitudes of the Fourier coefficients produced by Equation (2.2.5) are displayed in Figure 2.3. In order to fully represent **x** in the frequency domain, and to recover the original signal with Equation (2.2.8), it is necessary to sustain complex elements in the transform space. The *Inverse Discrete Fourier Transform* (IDFT) is defined as:

$$\text{IDFT} : \mathbb{C}^N \to \mathbb{C}^N \qquad (2.2.6)$$

now **x** is the image:

$$\mathbf{x} := \{x[n]\}_{n=0}^{N-1}, \quad x[n] \in \mathbb{C} \qquad (2.2.7)$$

where elements of **x** are calculated as:

$$x[n] = \frac{1}{N} \sum_{k=0}^{N-1} X[k] \cdot e^{i2\pi \frac{kn}{N}} \quad \text{for} \quad n = 0, 1, \ldots, N-1 \qquad (2.2.8)$$

By only looking at Figure 2.2, it is quite unintuitive to deconstruct the frequencies that went into this signal. As frequencies carry the recognizable information in signal data, having a method of wavelength decomposition is pivotal for data transfer.

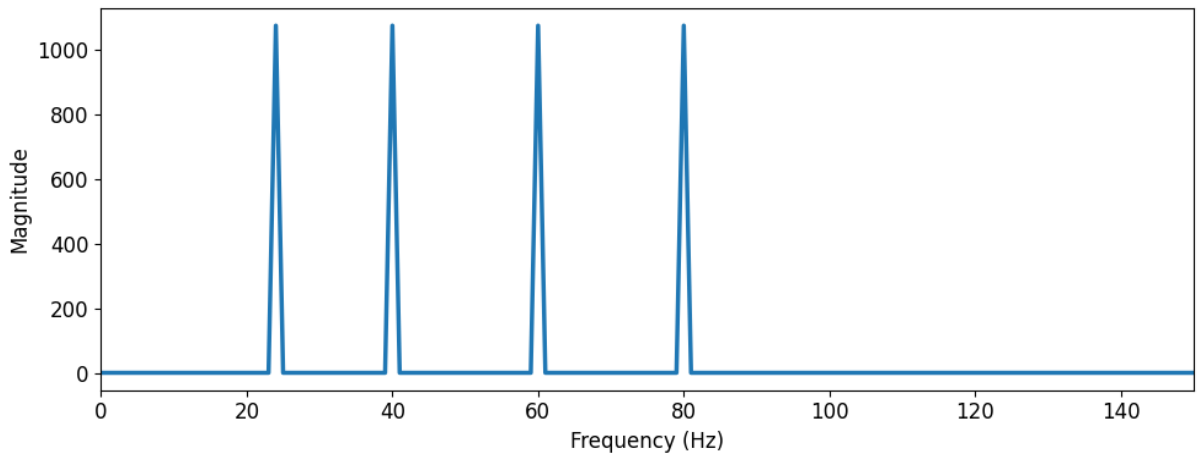

Figure 2.3: Fourier transform **X**.

Figure 2.3 highlights the original oscillators summed to produce the wave **x**. We easily note the magnitude peaks at 24, 40, 60, and 80 Hz. The times at which these frequencies appear are trivial because signal **x** is a stationary signal, meaning the frequencies are consistent over time. This class of signal would manifest in music as a consistent tone or set of tones, unhelpful for representing any real musical piece. For a more realistic example of a signal containing time-transient wavelengths, examine signal **y** being shown in Figure 2.4.

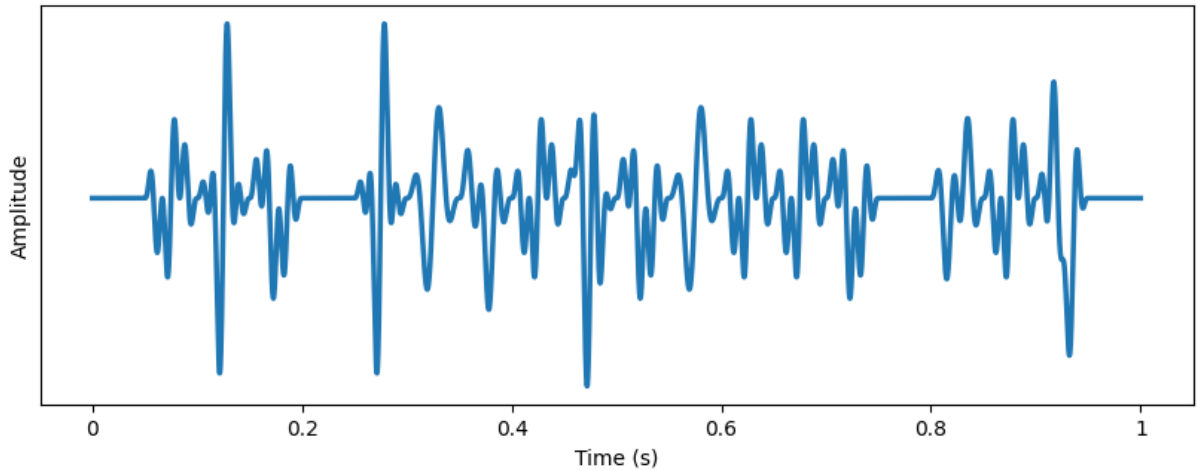

Figure 2.4: Transient frequencies in signal **y**

To show the frequencies contained in **y**, we apply Equation (2.2.5) to get **Y**: the transient signal's Fourier transform, as is shown in Figure 2.5. In this case, the overlapping frequencies that contributed to signal **x** of 40, 80, and 100 Hz are observed to be manifested in signal **y**. Due to the frequencies coming in and out randomly, the domain is noisy and there is little information we can reap from the barren soil of this time-independent representation. The times at which these frequencies are present in the signal is vital to its identity, introducing the motivation for a time-dependent frequency decomposition of **y**. This compromise analysis will be discussed in Section 2.4.

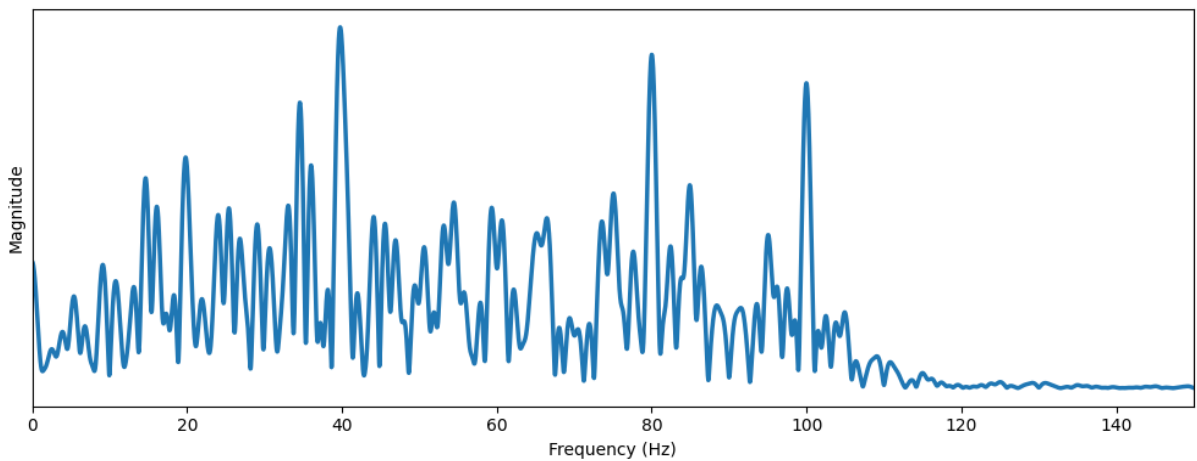

Figure 2.5: Fourier Transform **Y**

## 2.3 Sound, Music, and Frequency Analysis

When we experience a sound wave, our eardrum vibrates to match its frequencies, though we do not sense the individual oscillations of the drum. Rather, we have a familiar idea of how these frequencies register as pitches. A musical piece is easily defined as a sequence of frequencies, and this definition is quite sensitive to perturbations in both frequency and time. That is, a small change in frequency can result in a piece being unrecognizable, and similarly, small changes in time will unpleasantly contort the rhythm. This makes us perfect signal decomposers in both time and frequency domains.

Music exists in nature as a signal function, and it exists in digital storage as time series data. Neither of these representations are remotely recognizable as a musical piece— we are not able to distinguish the musical familiarity by looking at its sound wave.

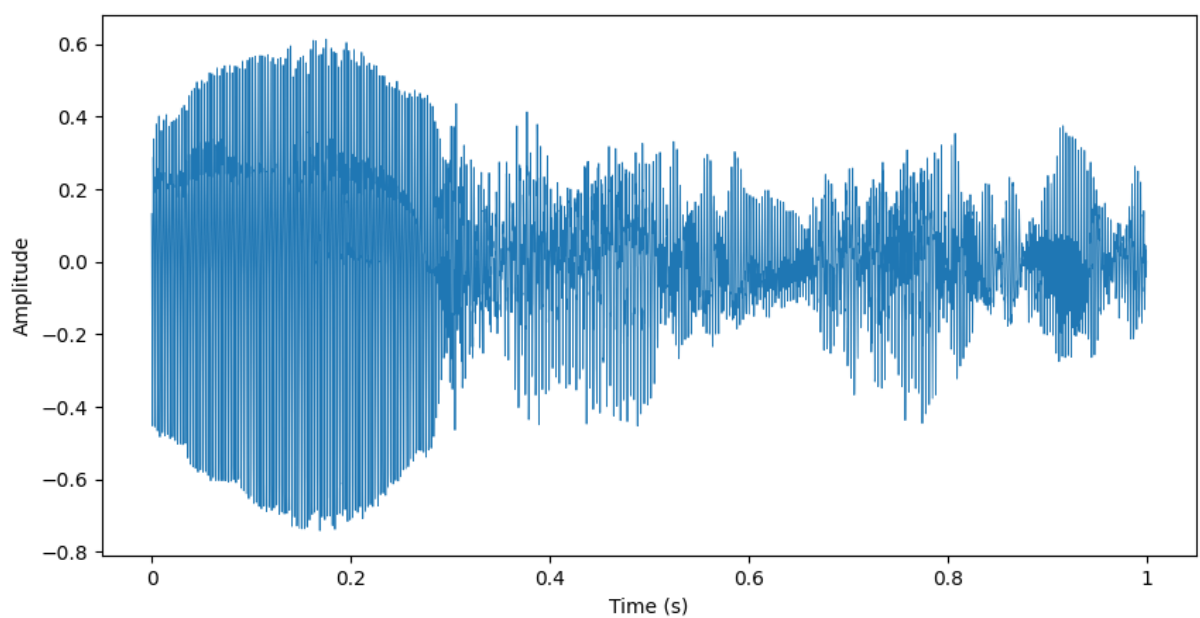

Figure 2.6: Audio waveform $\phi$.

Apart from crude metrics such as volume, we have no idea what was recorded to generate the waveform in Figure 2.6, which henceforth will be referred to as $\phi$. The astute among us could observe that there are multiple frequencies manifesting at each time, rather than simply being clean oscillations, which could be a clue that it was made with some instrument that generates harmonics. This will be discussed in further detail in Chapter 4.

Of course, this is only one representation of the sound wave, the one that exists in the *time domain*. We could also look at the very same wave in the *frequency domain*, as shown in Figure 2.7. From this perspective, all temporal information has been sacrificed, such as when the pitch is higher or lower or how the magnitude changes over time. To parallel the quantum wavefunction being viewed in position and momentum domains, these are dual representations of the same wave, connected by the Fourier transform.

Note that the Fourier coefficients range from a frequency of 0 Hz up to 4000 Hz. This is the maximum frequency that can be observed in this signal, being exactly half of the original sampling rate. This is known as the *Nyquist* frequency [5] which is commonly referred to as the upper limit of frequencies that can be observed in a signal. As a corollary, most digital audio is recorded at 44.1 kHz, implying a Nyquist number of 22.05 kHz. This recording technique operates under the assumption that 22050 Hz is the maximum listening frequency that a human might be interested in. From *An Introduction to the Psychology of Hearing*: "The range of human hearing extends from about 20 Hz to 20,000 Hz in young, healthy individuals, although the upper limit decreases with age."[6] By this standard, 44.1 kHz operates under a conservative estimate.

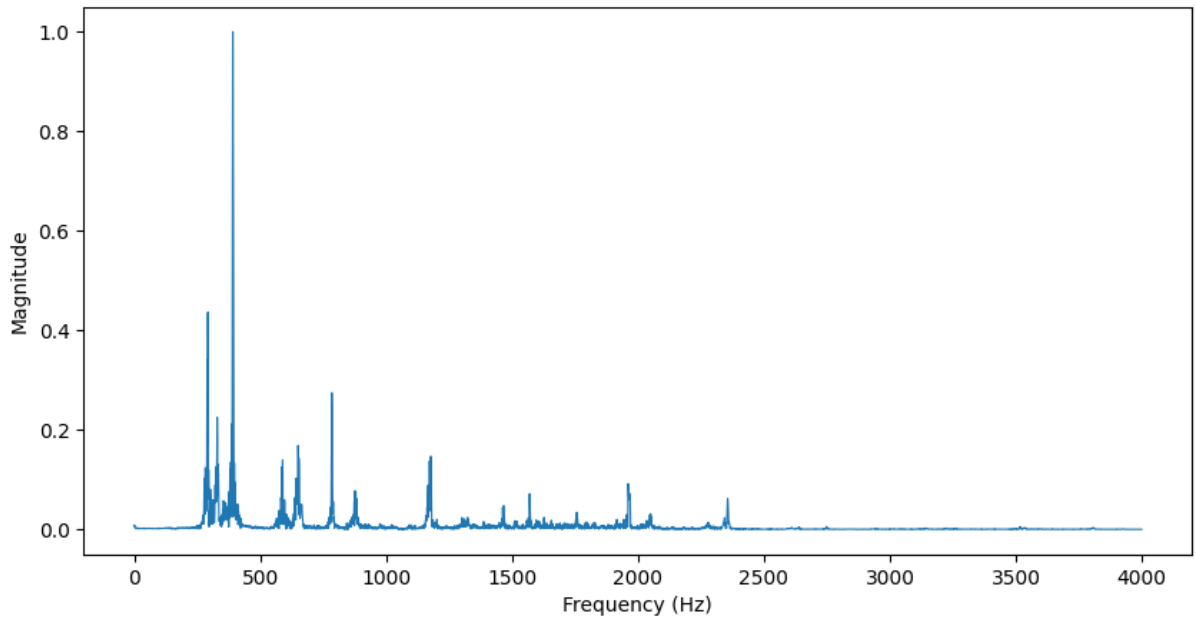

Figure 2.7: Magnitudes of the Fourier coefficients

While there is much that can be extrapolated from viewing the coefficient magnitudes on this spectra graph, we face the same limitations as we did in analyzing **Y** in Figure 2.5. If we know we are looking at a piece of music, one could guess not just the key or mode, but how heavily this piece relies on the root notes, versus branching off to more dissonant sounds. Although tangible information is intuitively read from this representation, neither Figure 2.3 nor 2.7 legibly depicts any specific features that reliably distinguish music. For these features to be recognizable, a balance between the two representations is required.

## 2.4 Time-Frequency Decompositions of Sound

In the interest of procuring an intuitive representation of our wave in Figure 2.6 ($\phi$), let us turn our attention to another picture, known as a *spectrogram*, in which we project the amplitude into a third dimension and focus on time and frequency information. We can begin by convolving the Fourier coefficients across our signal, creating an STFT as mentioned in Section 2.1. Because we are dealing with digital audio data, a discrete sum must replace the integral shown in Equation (2.1.7)[1]

$$Gabor_x[t, \omega] = \sum_{\tau=-\infty}^{\infty} x[\tau] \cdot g[\tau - t] \cdot e^{-i\omega\tau} \qquad (2.4.1)$$

And for our kernel, the vague window function $w$ in Equation (2.1.7) now gets defined by a Gaussian bell curve window (g), as shown in Equation (2.4.2). The replacement makes this STFT a *Gabor Transform* [9].

$$g[n] = e^{-\frac{1}{2}\left(\frac{n-N/2}{\sigma}\right)^2}, \quad 0 \leq n < N \qquad (2.4.2)$$

Where $n \in \{0, 1, \ldots, N-1\}$, $N$ is the window length parameter, and $\sigma$ represents the standard deviation, or *spread*, of the bell curve.

To understand how these Gaussian parameters affect the Gabor transform, consider the illustration of a singular time step shown in Figure 2.8. The highlighted time region shows where the kernel is active: the time step in which the Fourier coefficients are recorded. In this example, a wide window of $N = 1500$ samples is shown, with a standard deviation $\sigma = \frac{N}{6}$. If the audio of signal **x** is sampled at 8000 Hz, this is equivalent to a time window of about 0.19 seconds. Recall that, according to Equation (2.1.6), a wider window produces higher accuracy in the frequency domain while sacrificing temporal precision.

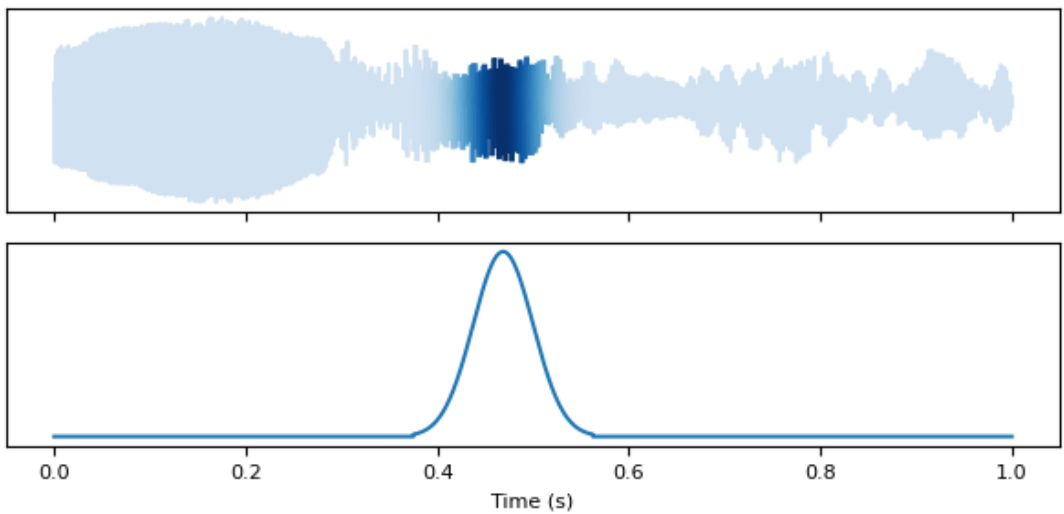

Figure 2.8: Gaussian window at one time step in signal **x**

---

[1] In practice, $x[\tau]$ is finite-length and assumed zero outside its support, so the integral or sum is taken over the effective signal window. We write the $[-\infty, \infty]$ boundaries here as a formality, to indicate this as a discretized convolution. The same logic applies to Equation (2.1.7).

Substituting $\phi$ in for the $x[\tau]$ in Equation (2.4.1) produces the spectrogram shown in Figure 2.9, with time displayed on the x-axis, frequency on the y-axis, and the amplitude is represented by color. This heatmap provides a superior representation of the musical information of $\phi$ because all three of these dimensions are crucial to the identification of a musical fingerprint. As a first look at the time-spectra of $\phi$, 2.9 was generated using a conservatively small $\sigma$ value.

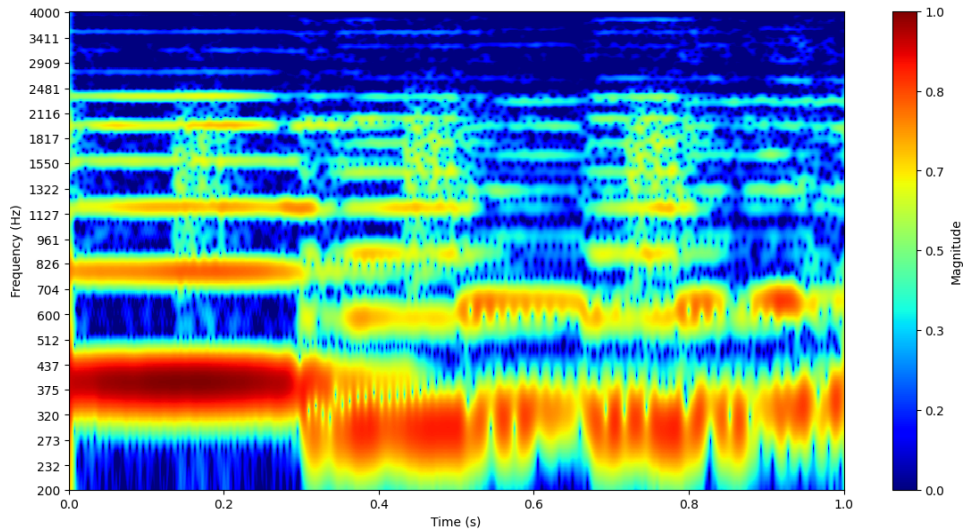

Figure 2.9: Gabor transform of signal $\phi$

The Gaussian used to produce Figure 2.9 was heavily weighted for temporal accuracy. Notice that we have significant vertical bleeding, to the point where the original frequencies to be captured are barely recognizable. Any sound containing such a broad consecutive spectrum of frequencies would simply manifest as white noise. This limited precision on the y axis is more apparent when held against a Gabor transform yielded from a high spread Gaussian kernel, as seen in Figure 2.10. In this case we observe horizontal overlap of the dark regions, as if they've been smeared like paint across the array. However, they do maintain clean delineations along the vertical.

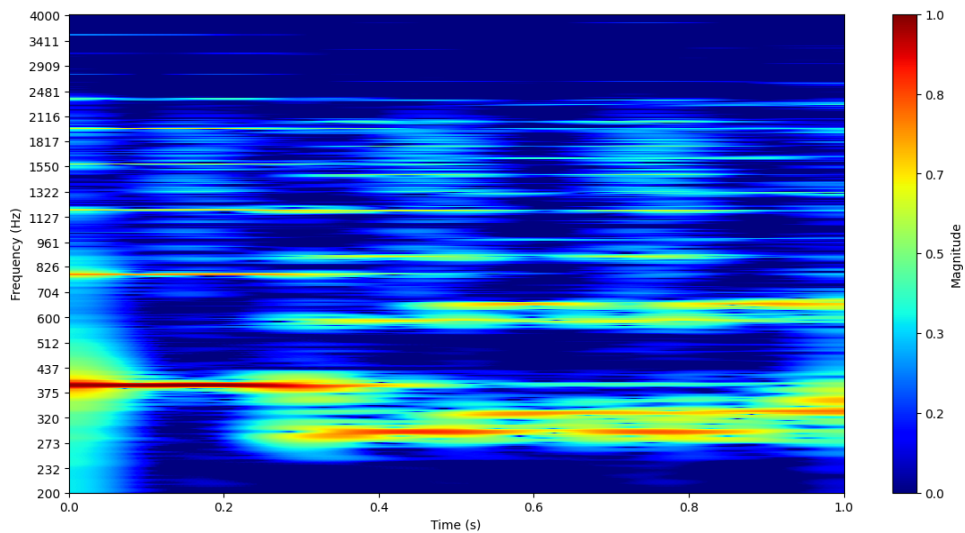

Figure 2.10: Gabor transform with wide kernel

If the mission is to find a musically intuitive representation of $\phi$, the Figure 2.10 is an over-correction of the problem observed in Figure 2.9. To find a balance between the two, we adjust $\sigma$ until we are satisfied in distinguishing the frequencies while also having a solid estimate of when they are heard. This *Goldilocks* transform is shown in Figure 2.11. A clear tone right around 390 Hz is seen at the beginning, sustaining for approximately 0.3 seconds, after which it has about a 0.1 second release. Note that, for Figures 2.9, 2.10, and 2.11, the y-axis is scaled logarithmically. This is a purely cosmetic choice, meant to emphasize those frequencies in which most music is explicitly written, typically ranging from 200 to 1000 Hz.

The balanced Gabor transform in Figure 2.11 is a useful compromise between time and frequency resolution, allowing us to distinguish discrete pitches and temporal onsets with sufficient clarity to trace musical events — such as sustained tones, note attacks, and decay characteristics. Although this is a more familiar representation of our signal, there are significant limits to our comprehension of the static time-frequency mesh. In the following section, we will explore algorithmic and perceptual heuristics for problem and see computationally viable, yet expensive solutions to it.

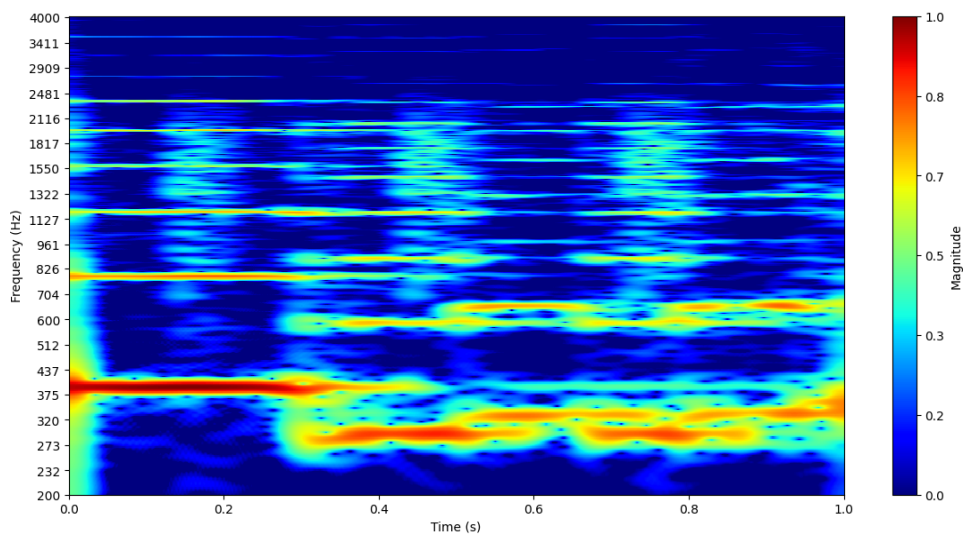

Figure 2.11: Gabor transform with balanced kernel

## 2.5 The Continuous Wavelet Transform

After laying out the problems with trading certainty between time and frequency, we should explore a dynamic approach to time-frequency analysis with the *wavelet transform*. For this method, instead of taking independent frequency spectra within short windows, we convolve a localized wave function, known as a wavelet, against our signal, producing magnitudes of representation for each timestep of the convolution. An example of a wavelet function, the *Morlet* wavelet [8], is shown in Figure 2.12, where we see the real and imaginary parts plotted independently.

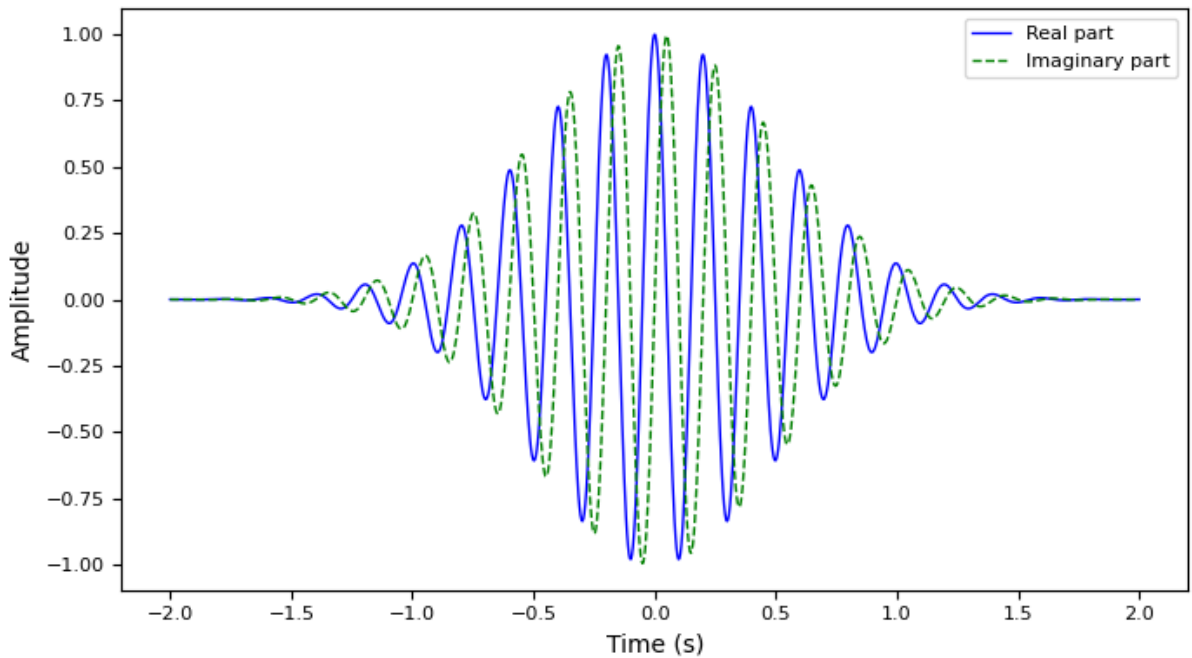

Figure 2.12: Morlet wavelet with $f_0 = 5$.

Just as the classical Fourier transform gave us the phase information of the signal as complex numbers, so too can the wavelet transform if employing a complex-valued wavelet such as the Morlet. If we are interested only in the frequencies and times at which they occur, as is the case with music, this phase information is simply a by-product, though broad-domain applications that utilize the phases are discussed in Chapter 5.

The wavelet transform operates by scaling the wavelength of a *mother wavelet* across whichever band of frequencies we are interested in. Now, rather than exchanging accuracy between our domains of interest, we can choose whatever resolution in frequency we desire by adding more scaled wavelets. The only limitation in time resolution comes from the uncertainty principle applied individually at every scale. This means that our time precision improves as the wavelength of the wavelets decreases. In contrast with Figure 2.1, the dynamic mesh of the wavelet transform can be seen in Figure 2.13. Each blue stripe represents one scaled wavelet's convolution across the signal, and as the scale increases, we see the stripes have better resolution in time, but worse in frequency.

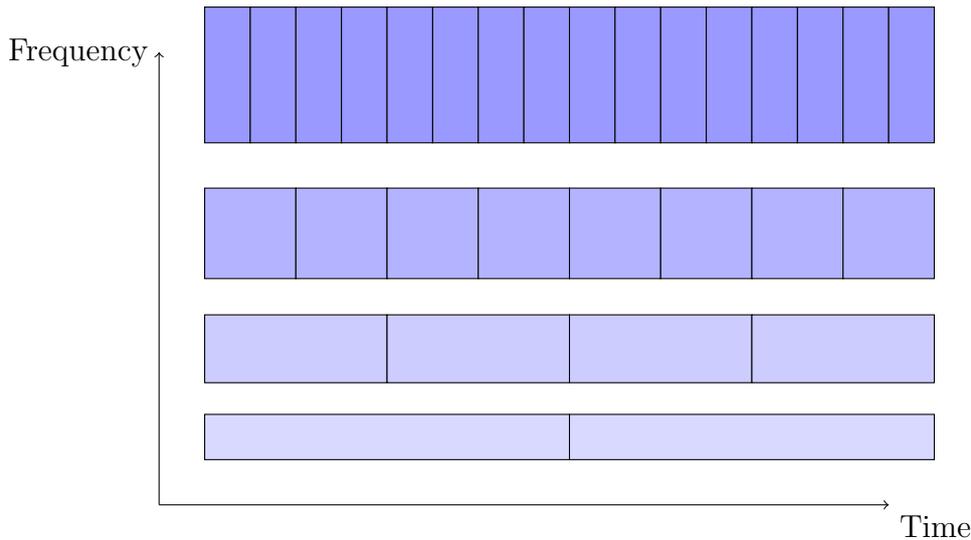

Figure 2.13: Wavelet transform mesh

This phenomenon, which applies to all waves being measured, occurs because less time is required to observe the gap between peaks when the wavelength is low. To illustrate this principle with an example, let us look again at signal **y**. The real component of a Morlet wavelet at three different scales is active at a singular timestep in **y** in Figure 2.14, each scale preserving the same number of oscillations. Thus, the smaller wavelength takes up a smaller window and we get a finer, more localized view of the frequencies at this timestep.

These scaled and shifted versions of the mother wavelet ($\psi$) are formalized in the *Continuous Wavelet Transform* (CWT). For a signal $x(t)$, the CWT is defined in Equation (2.5.1).

$$\mathcal{W}_x(a,b) = \frac{1}{\sqrt{|a|}} \int_{-\infty}^{\infty} x(t) \cdot \psi^* \left( \frac{t-b}{a} \right) dt \tag{2.5.1}$$

Where:

- $a$ is the *scale* parameter, inversely related to frequency

- $b$ is the *translation* (or time shift)

- $\psi^*$ is the complex conjugate of the mother wavelet

It is important to note that other mother wavelets can be substituted in for $\psi$, although the Morlet is most commonly associated with the continuous wavelet transform [7]. We define the Morlet as:

$$\psi(t) := \pi^{-\frac{1}{4}} e^{if_0 t} e^{-\frac{t^2}{2}} \tag{2.5.2}$$

This wavelet is essentially a complex sinusoid modulated by a Gaussian envelope, bearing strong resemblance to the Gabor kernel discussed earlier. The key parameter here is $f_0$, the central frequency of the wavelet, around which the other frequencies in the wavelet transform are scaled.

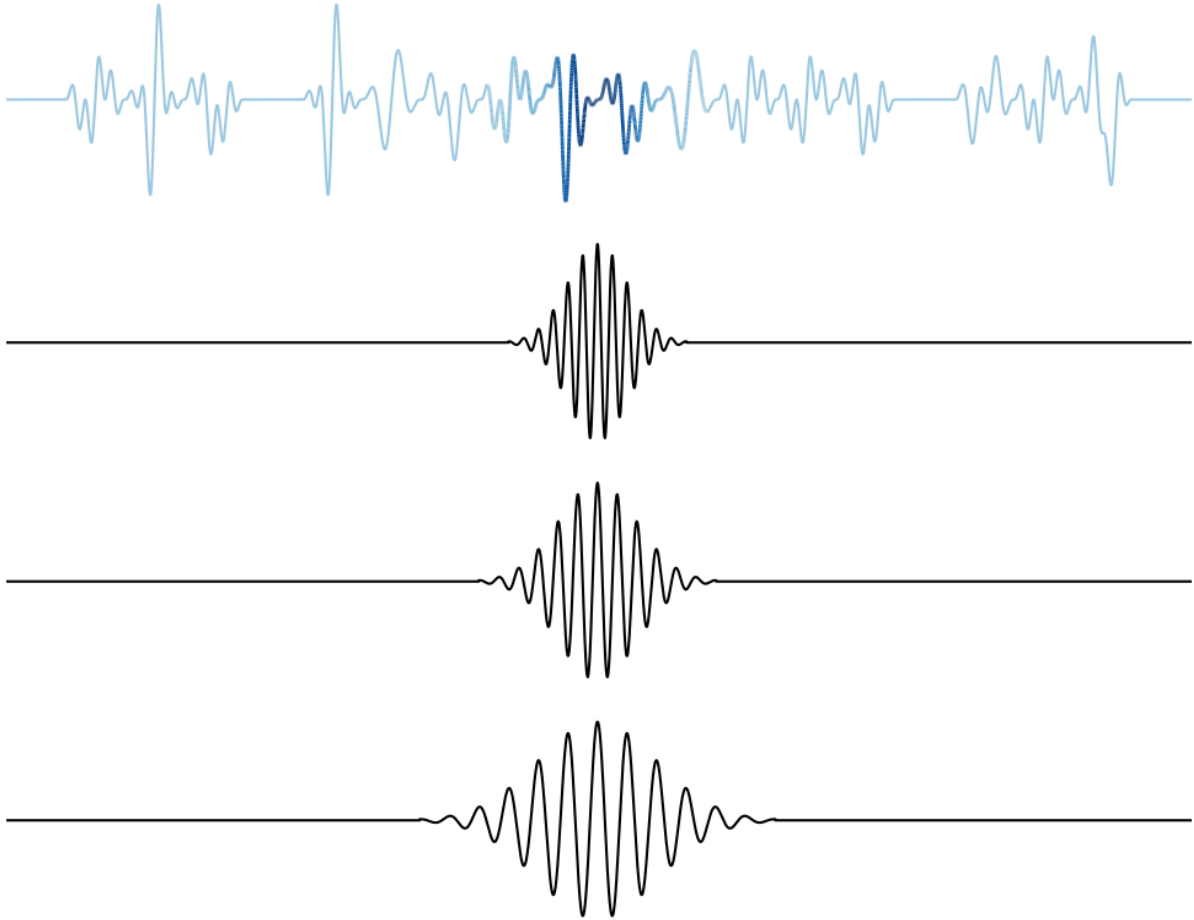

Figure 2.14: Three scaled Morlets at a timestep in **y**

The full convolutions in time for each of the three wavelets in Figure 2.14 are displayed in Figure 2.15. This would be the result if we performed Equation (2.5.1) on signal **y** at only three scale values of $a$, those corresponding to 40, 80, and 100 Hz. Note how the resolution in time increases as the wavelength decreases, in accordance with the meshing in Figure 2.13. Recall also that these are complex oscillators, so only the real components are seen in Figure 2.15. The magnitudes of the complex wavelet coefficients are plotted in the three-banded proto-spectrogram in Figure 2.16[2].

---

[2]Figure 2.16, and all spectrograms shown moving forward, do not include numerical labels on the **Magnitude** colorbar. While we can assume these magnitudes range between 0 and 1, as shown in Figures 2.9, 2.10, and 2.11, the absolute values are in fact irrelevant, as we are only concerned with the relevant magnitudes within the time-frequency space. Similarly, all sound waves shown moving forward do not include labels on the amplitude (y) axis, as we do not care about the overall volume of a sound, just the relative volume between temporal locations

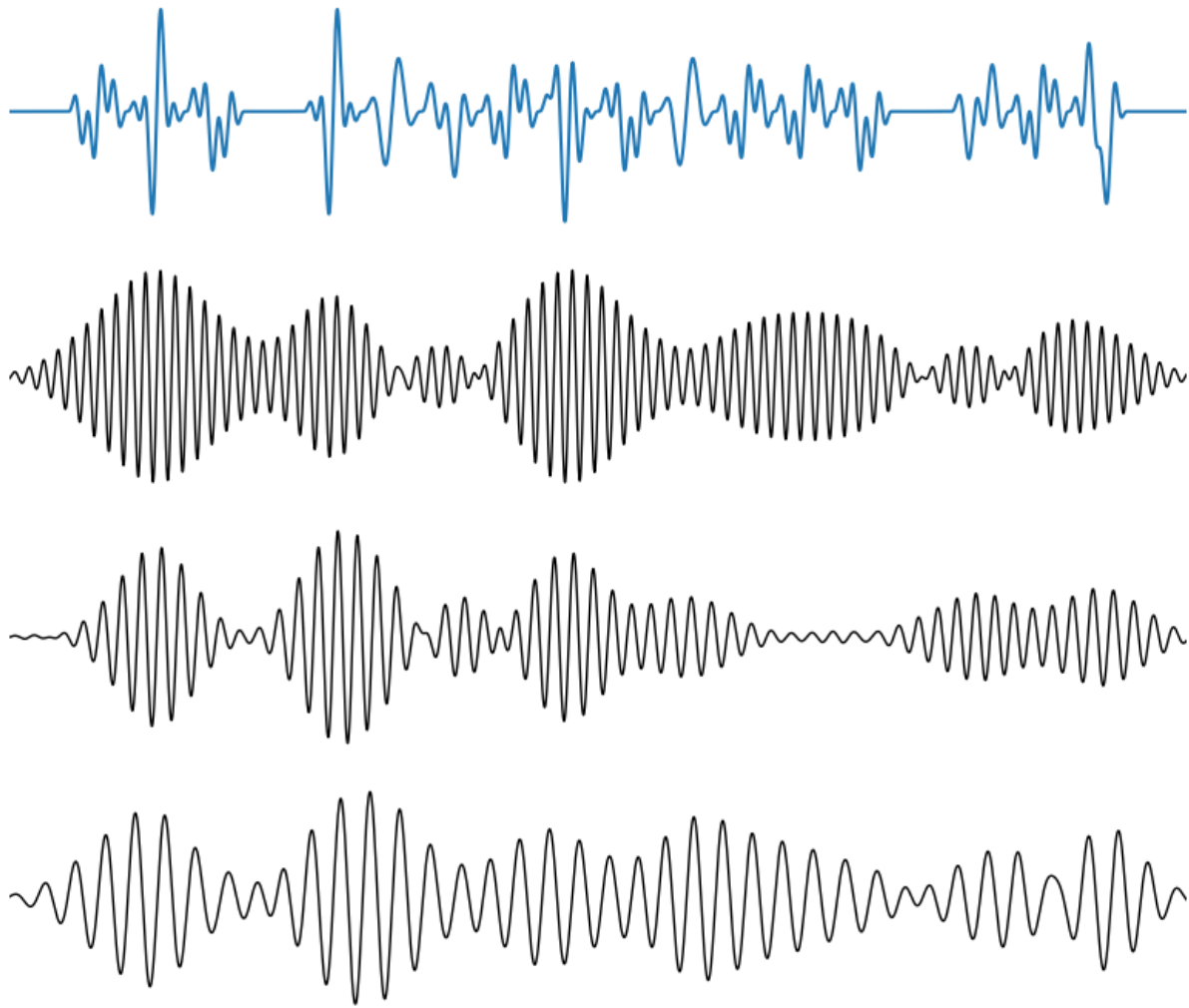

Figure 2.15: Real parts of the wavelet coefficients of **y** at three scales.

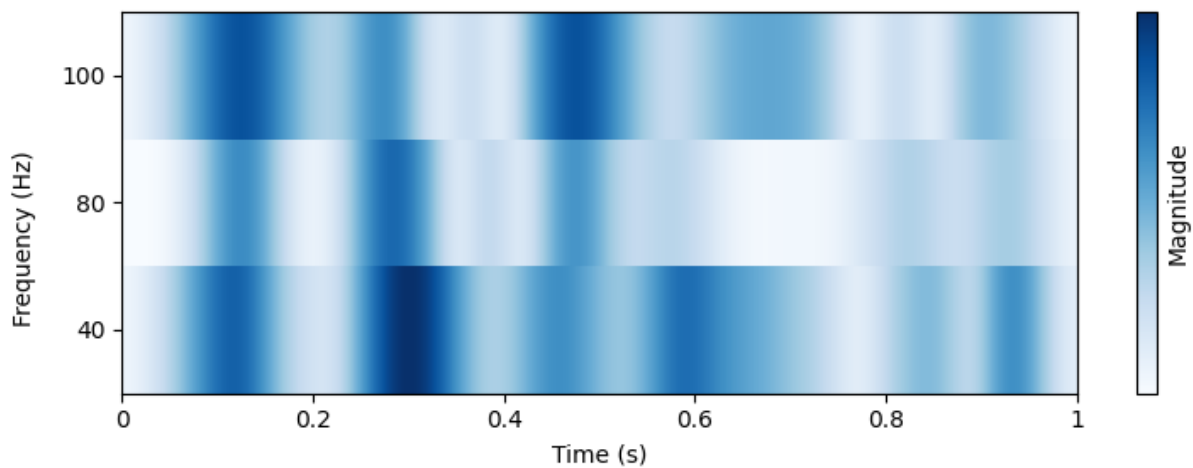

Figure 2.16: Magnitudes of wavelet coefficients for three scales on **y**.

To wrap this discussion back into the problem of spectrogram time-frequency resolution from Section 2.4, we now apply the continuous wavelet transform to the same signal $\phi$ from Figure 2.6. The result, visualized in Figure 2.17, provides a multi-resolution view of the signal. Compared to the Gabor transform in Figure 2.11, the CWT is able to preserve detail across a broader range of frequencies, while capturing sharp onsets and releases with high fidelity.

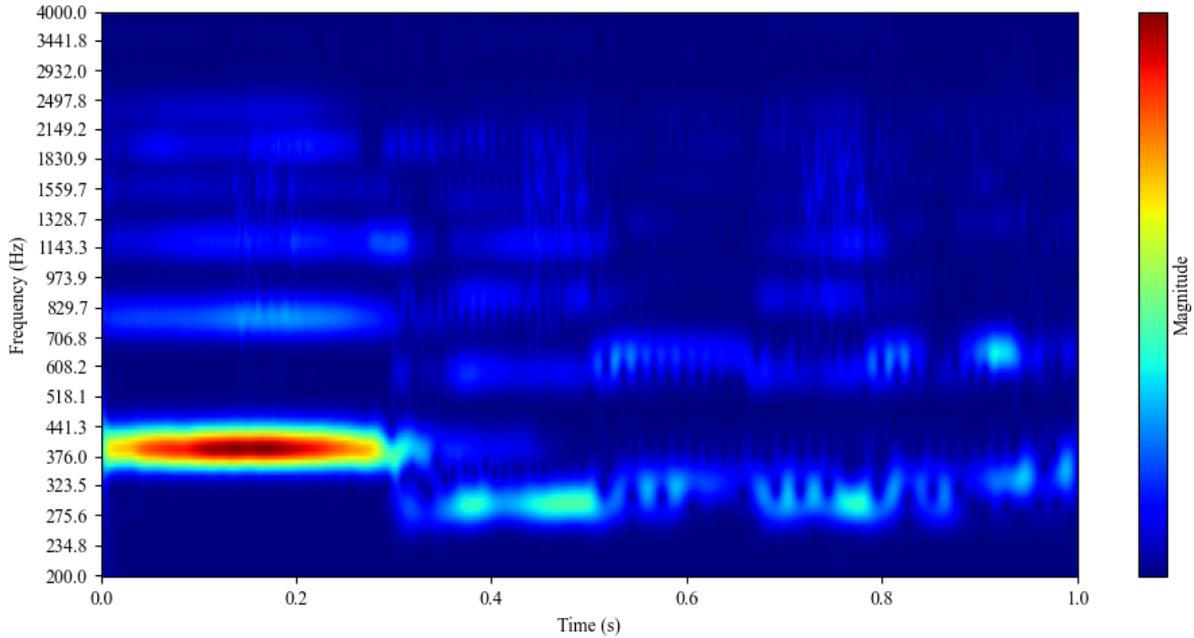

Figure 2.17: Continuous Wavelet Transform of signal $\phi$

This flexibility makes wavelet transforms especially useful for pattern recognition and feature detection in music. Though computationally more expensive than STFT, the CWT's superior resolution trade-offs make it a powerful foundation for subsequent feature extraction and machine learning tasks, which will be discussed in Chapter 5. The following chapters will discuss how we can use these methods directly, in conjunction with *Wavelet Coherence Analysis*, to categorize and identify Irish music.

# Chapter 3

# Methodology

By introducing an application for the wavelet transform to traditional Irish music, we prove the capabilities of the wavelet transform and wavelet coherence models.

## 3.1 Data Description

For this application, we have two types of data to feed into our models:

- Live audio recordings

- Musical Score Notation

For the audio, all inputs are recorded at 8000 Hz, and for exactly 19.20 seconds. This makes each audio input a time series vector of length $n = 153.6 \times 10^3$ samples. An example of this data type can be seen in Figure 3.1.

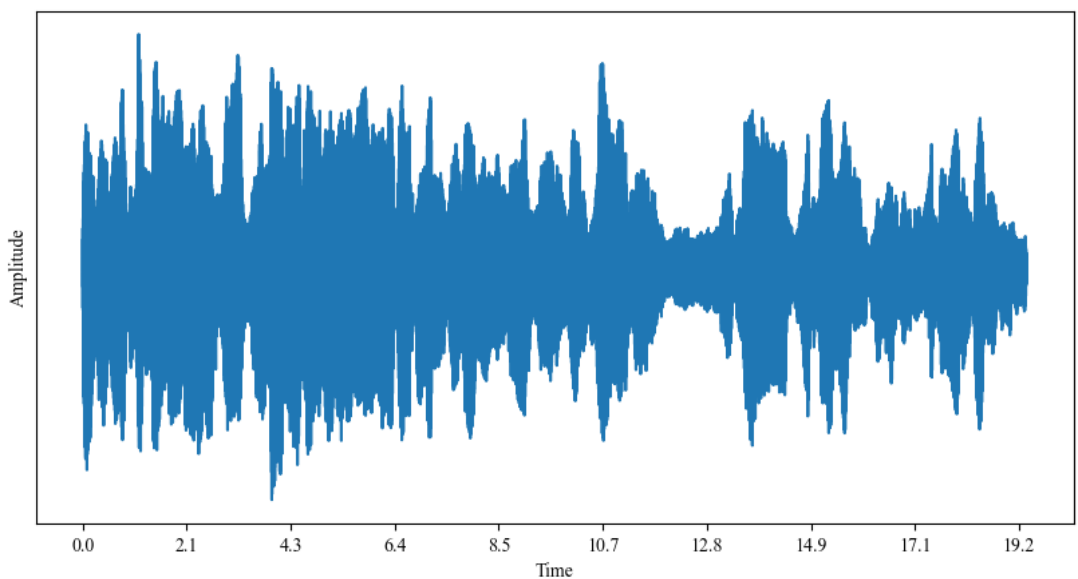

Figure 3.1: 19.20 seconds of audio at 8000 Hz.

These particular dimensions are downstream of choices made about how to compare the recordings with the sheet music data. In this research we are only looking at reels, and have made some aforementioned edits to ensure their equal length. We are choosing to record everything at exactly 100 BPM, and we are aiming to record only reels containing exactly 32 beats. This rate of 0.6 seconds per beat ensures that each note has enough sustain to be recognized on our spectrogram.

For the sheet music data, all scores come from contributors on *The Session* [25], and we are using a slightly padded version of the list curated by Mc Gettrick et al. [13]. This list includes 78 reels, and Figure 3.2 is one such example.

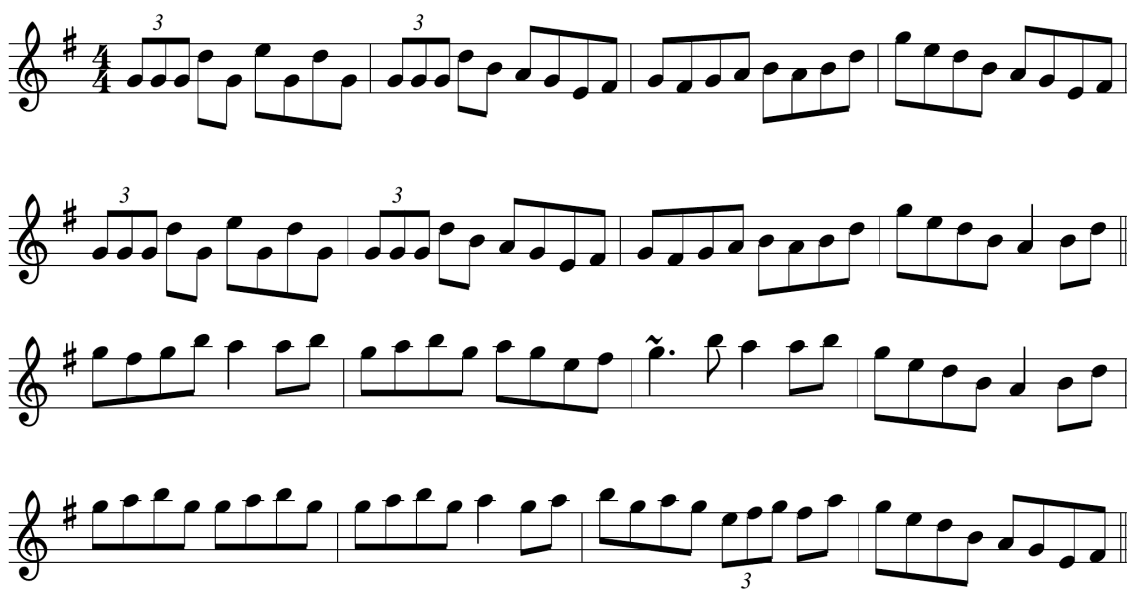

Figure 3.2: *The Galway Rambler* score.

Section 3.2 will focus on converting this annotated format into a form that can be compared with the recording.

## 3.2 Symbols and Waveforms

To make an apples-to-apples comparison between the symbolic representation of music and the raw sound, we need to generate a synthetic waveform that matches what is written in Figure 3.2. We can begin by retrieving the tune's ABC notation, which is saved in the following format:

```
X: 1
T: The Galway Rambler
R: reel
M: 4/4
L: 1/8
K: Gmaj
(3GGG dG eGdG|(3GGG dB AGEF|GFGA BABd|gedB AGEF|
(3GGG dG eGdG|(3GGG dB AGEF|GFGA BABd|gedB A2 Bd||
gfgb a2ab|gabg agef| g3 b a2ab|gedB A2 Bd|
gabg gabg|gabg a2ga|bgag (3efg fa|gedB AGEF||
```

For each tune listed on The Session, there are multiple *settings* (transcribed versions) uploaded, one of which is displayed above for The Galway Rambler. The `X: 1` indicates that this setting has an index of 1. Other metadata in preamble lists the *title, dance, meter, length* and *key*. Additionally, the lack of ||: and :|| symbols bookending the first and second halves of the ABCs indicate that this is a **single reel.**

This tune contains a total of 16 bars at 4/4 time, with each bar containing 4 quarter notes or 8 eighth notes. If we divide the tune into 128 time slots, we can represent it by filling each slot with a letter. Using this method, we have to treat each point on the grid independently, i.e. all notes are treated as eighth notes. The differences in meter are dealt with in the following ways:

- The long notes, such as quarters, dotted quarters, or half notes, are broken into individual eighth notes.
    - g3 ⟶ ggg,   A2 ⟶ AA

- Triplets kick out the middle note, treating the first and third as eighth notes.
    - (3GGG ⟶ GG,   (3efg ⟶ eg

To properly read this, the *key* is vital as well. A signature of G major signals to us that the `f` is sharpened, so we should indicate that with a $\hat{\texttt{f}}$. The different letter case also denotes the octave in which the note lies: `C` would indicate *middle C* on the piano, with a frequency of 261.63 Hz, while `c` is an octave above. With this, we are ready for the next step in our pipeline: encoding semitones from the ABC notation. We use the following encoder to represent the tune as a sequence of tones, each integer indicating the number of semitones above middle C.

Given a value $n$ in this sequence, the following formula delivers the corresponding frequency:

$$f = f_0 \cdot 2^{n/12} \qquad f_0 = 261.63 \, Hz \tag{3.2.1}$$

$$C \to 0 \quad \hat{C} \to 1 \quad D \to 2 \quad \hat{D} \to 3 \quad E \to 4 \quad F \to 5 \quad \hat{F} \to 6 \quad G \to 7$$
$$\hat{G} \to 8 \quad A \to 9 \quad \hat{A} \to 10 \quad B \to 11 \quad c \to 12 \quad \hat{c} \to 13 \quad d \to 14 \quad \hat{d} \to 15$$
$$e \to 16 \quad f \to 17 \quad \hat{f} \to 18 \quad g \to 19 \quad \hat{g} \to 20 \quad a \to 21 \quad \hat{a} \to 22 \quad b \to 23$$

For example, the `a` note above middle C would have $n = 9$ semitones, yielding:

$$f = 261.63 \cdot 2^{9/12} = 440 \, \text{Hz} \qquad (3.2.2)$$

Recall that our sequence is 128 tones long, and has a total of 32 beats. If we want to simulate *The Galway Rambler* at 100 BPM, we need each tone to last for exactly 0.15 seconds. At 8000 Hz sampling rate, each tone is constructed into a sine wave containing exactly 1200 samples. This will bring the total length of the constructed waveform to $153.6 \times 10^3$ samples, matching the length of the recorded audio from Section 3.1.

The first three notes of *The Galway Rambler* can be seen as their waveform rendering in Figure 3.3.

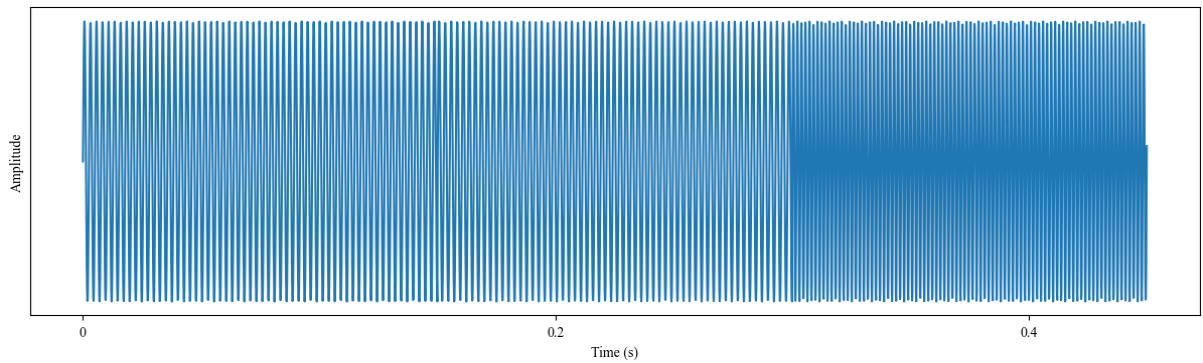

Figure 3.3: First half second of *The Galway Rambler* (synthesized).

Optionally, we can add some harmonics and give the waveforms individual envelopes to better simulate a real instrument, as seen in Figure 3.4. The wavelet transform for this synthesized clip can be seen in Figure 3.5.[1]

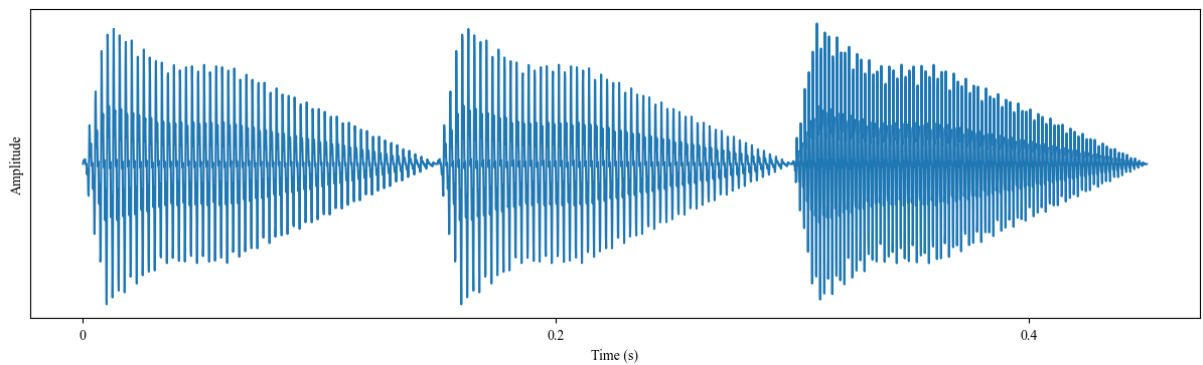

Figure 3.4: *The Galway Rambler* opening with harmonics and envelopes.

---

[1] The code containing the specific harmonics and envelopes can be found in Appendix A.

## 3.3 Wavelet Transform Implementation

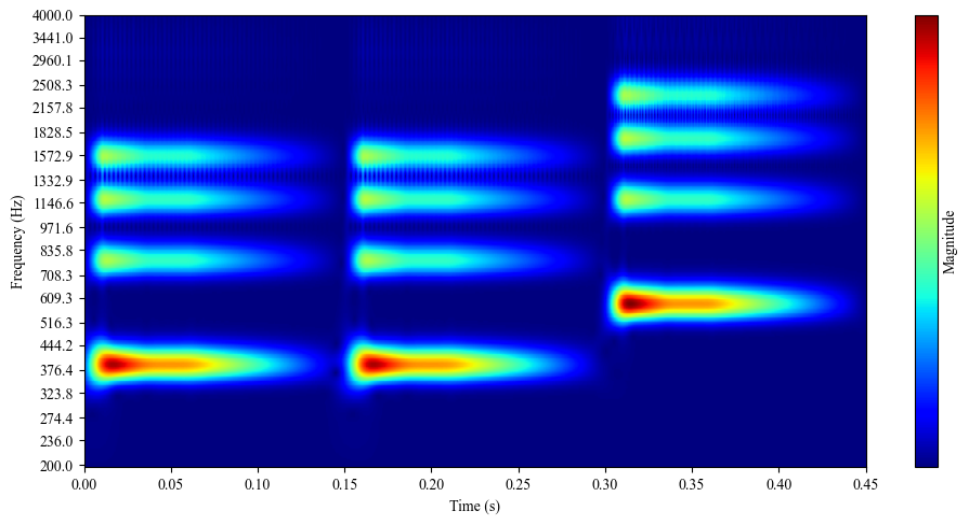

Figure 3.5: Example wavelet transform.

After our data has been properly sautéed and stitched up, we are ready to implement the CWT. Beginning with the recording, the time series data are fed into a numeric solver for the CWT integral. For efficiency, the wavelet coefficients produced by this model are computed using the *Fast Continuous Wavelet Transform* (fCWT), which allows us to make practical use of this expensive convolution. More information on fCWT will be discussed in Chapter 5.

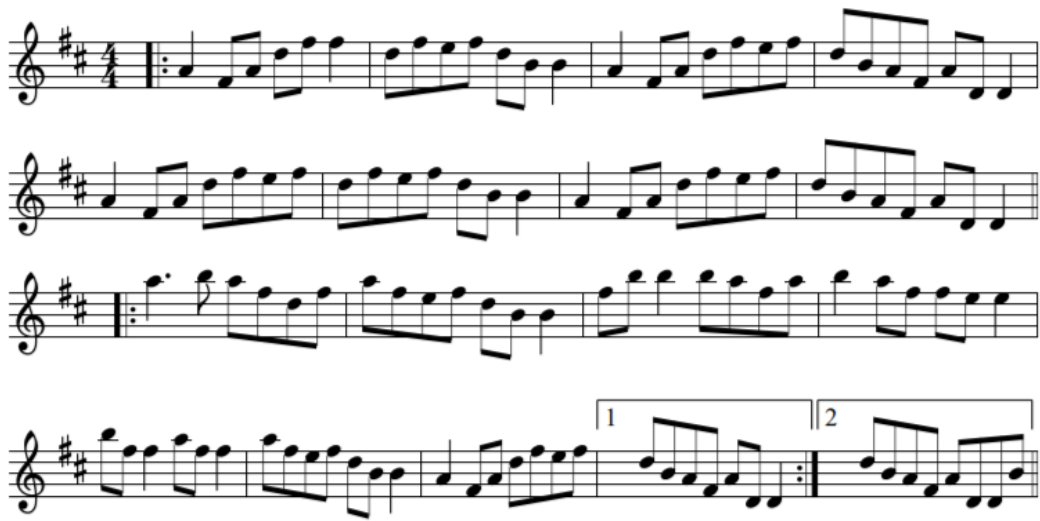

Figure 3.6: *The Sailor's Bonnet* score.

As an example, let us start with the written score for *The Sailor's Bonnet,* displayed in Figure 3.6 [2]. Using the pipeline outlined in the previous section, the synthetic spec-trogram is produced by the score, and can be seen in Figure 3.7. One can observe the harmonic overtones encoded into each note, as well as see how the long notes, like the first note in *The Sailor's Bonnet*, are broken up.

---
[2] The last measure, after the `:||` symbol, has been removed from the ABC notation in this work, as it is irrelevant.

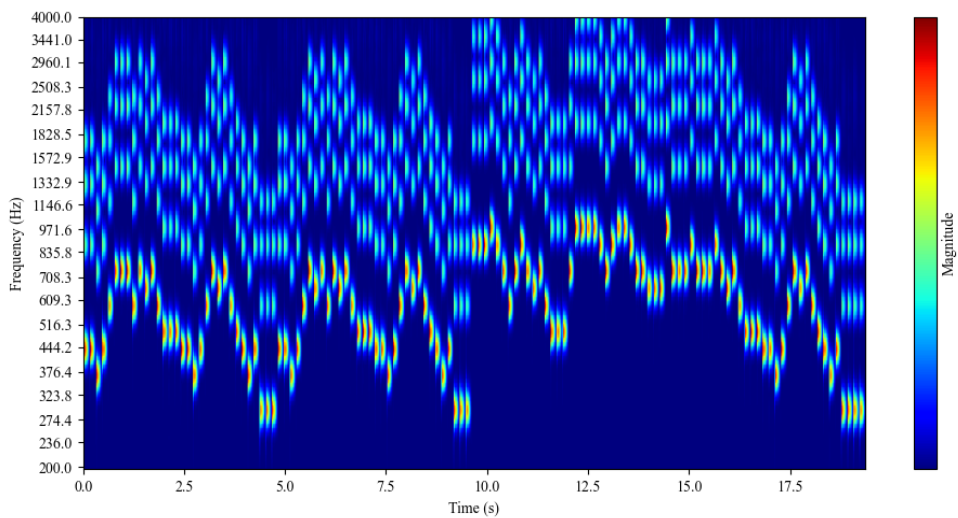

Figure 3.7: *The Sailor's Bonnet* Synthetic Spectrogram.

To ensure that the recording matches up with the synthetically generated waveform, this model has a built-in metronome and count-in. This is necessary to compare what is being played with the written score. Figure 3.8 shows the recorded waveform, as well as the computed wavelet transform.

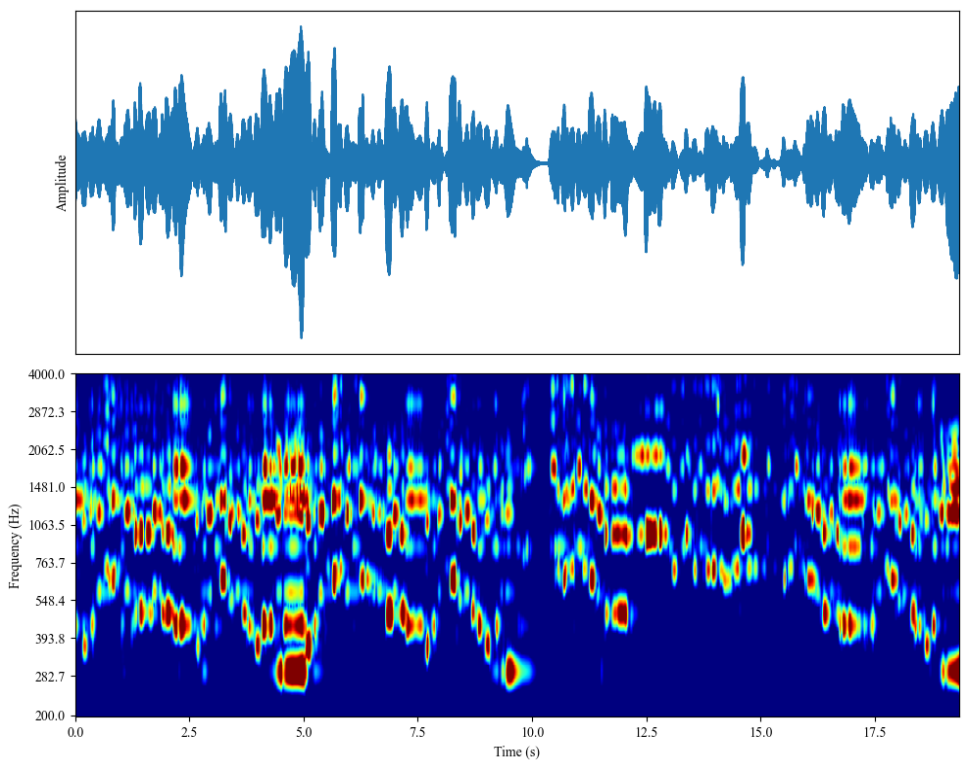

Figure 3.8: *The Sailor's Bonnet*, played on fiddle.

## 3.4 Wavelet Coherence and Pattern Matching

At long last we are at the crux of the tune identification example: *wavelet coherence analysis*. This powerful method is used across domains to study the relationships between time series data. If we have two signals, $x$ and $y$, the wavelet coherence is computed by the following formula:

$$\mathcal{C}(a,b) = \frac{|S\left(\mathcal{W}_{xy}(a,b)\right)|^2}{S\left(|\mathcal{W}_x(a,b)|^2\right) \cdot S\left(|\mathcal{W}_y(a,b)|^2\right)} \tag{3.4.1}$$

Where $S$ is a customizable smoothing operator, defined depending on how loose we allow our search for time-frequency relationships to be. Similar to Equation (2.5.1), $a$ and $b$ represent scale and time, respectively.

We must also define here the *cross-wavelet transform*, written in Equation (3.4.1) as $\mathcal{W}_{xy}$.

$$W_{xy}(a,b) := W_x(a,b) \cdot W_y^*(a,b) \tag{3.4.2}$$

As this is an element-wise operation between two arrays of complex numbers, $W_x(a,b)$ and $W_y(a,b)$, the multiplication between the first element and the complex conjugate of the second returns a complex number whose argument is the difference between phases. The benefit of this being that, for each frequency shared between $x$ and $y$, we can see the phase shift between the signals. Applications of this phase differential will be discussed in Chapter 5, though it will not be relevant for the musical example in this work.

As we have prepared two wavelet transforms in the previous section, we are ready to implement the wavelet coherence on our musical data. Letting $x(t)$ be the recorded waveform and $y(t)$ be the synthetic, we already have the spectrograms at our disposal. After substituting these arrays for $\mathcal{W}_x$ and $\mathcal{W}_y$, we get our matrix $\mathcal{C}$, which is displayed in Figure 3.9.

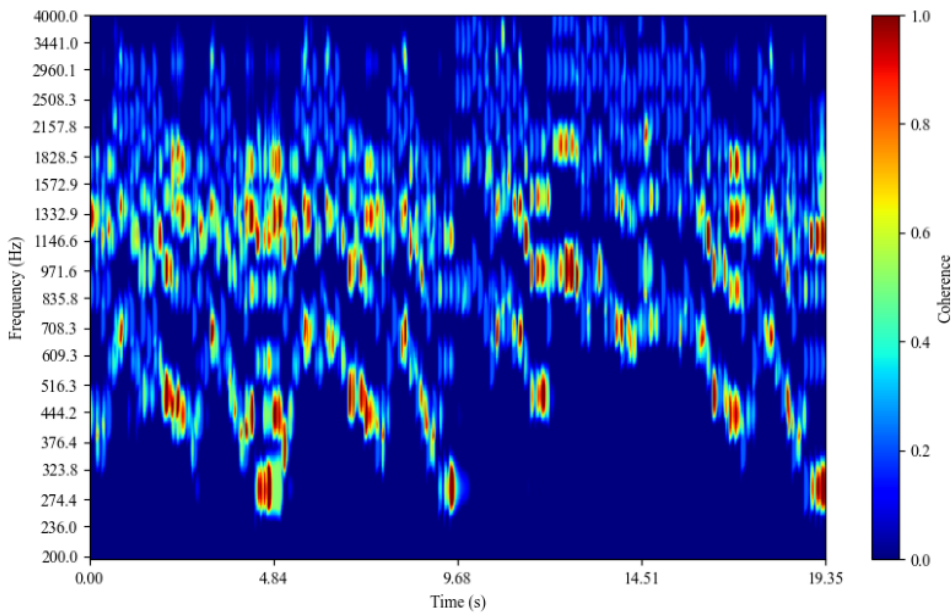

Figure 3.9: *The Sailor's Bonnet* Coherence

# Chapter 4

# Results

Using the outlined methods of the wavelet transforms and wavelet coherence, the tune recognition model's capabilities are demonstrated.

## 4.1 Spectrogram Fingerprinting

Figure 3.9 displays the coherence between two signals that are alternative representations of the same tune; one being symbolic, and the other being audio. To examine the utility of this, consider another coherence array in Figure 4.1, where the inputs are *not* representing the same tune. Here, the recording of *The Sailor's Bonnet* is matched against the score of *The Galway Rambler*. Note how the red regions are few and far between, especially compared with Figure 3.9.

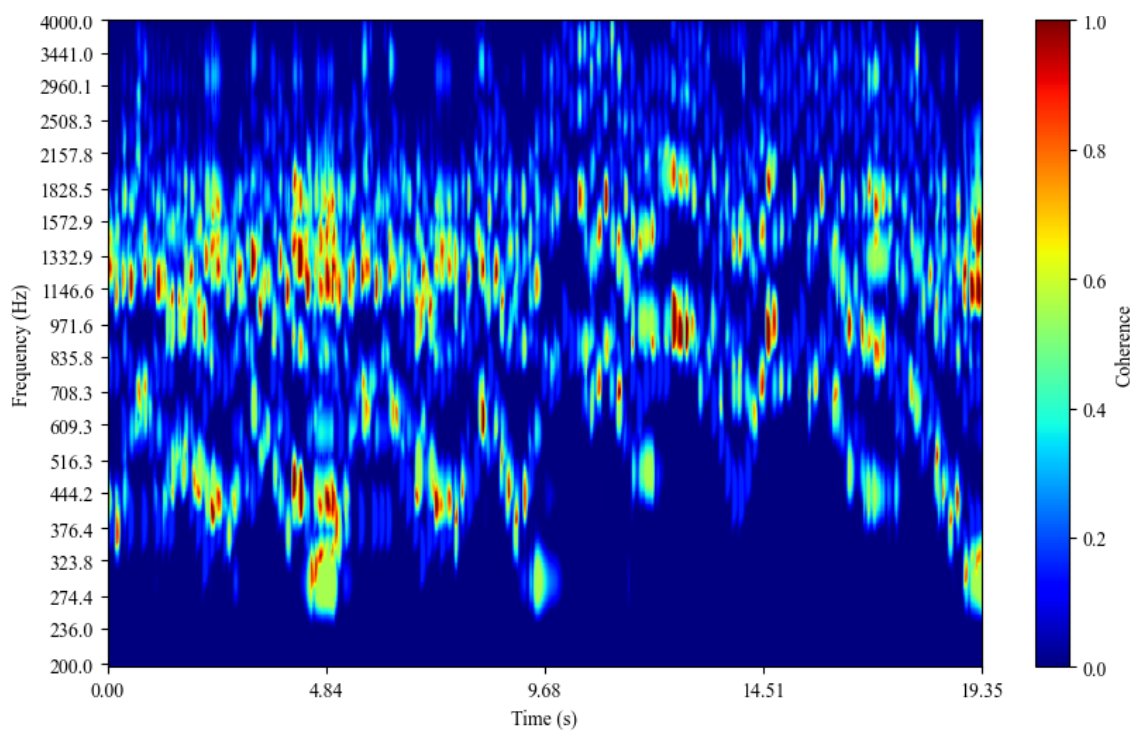

Figure 4.1: Mismatched coherence.

The key factor in the identification model is this head-to-head comparison between transforms. If we have the spectrogram of a recording matched against a database of tune scores, any crude metric could hint at what was recorded. For *The Sailor's Bonnet* recording of Figure 3.8, the total (gross) coherence of each pairing, produced by summing the entire array, in the tunebase is listed in Figure 4.2.

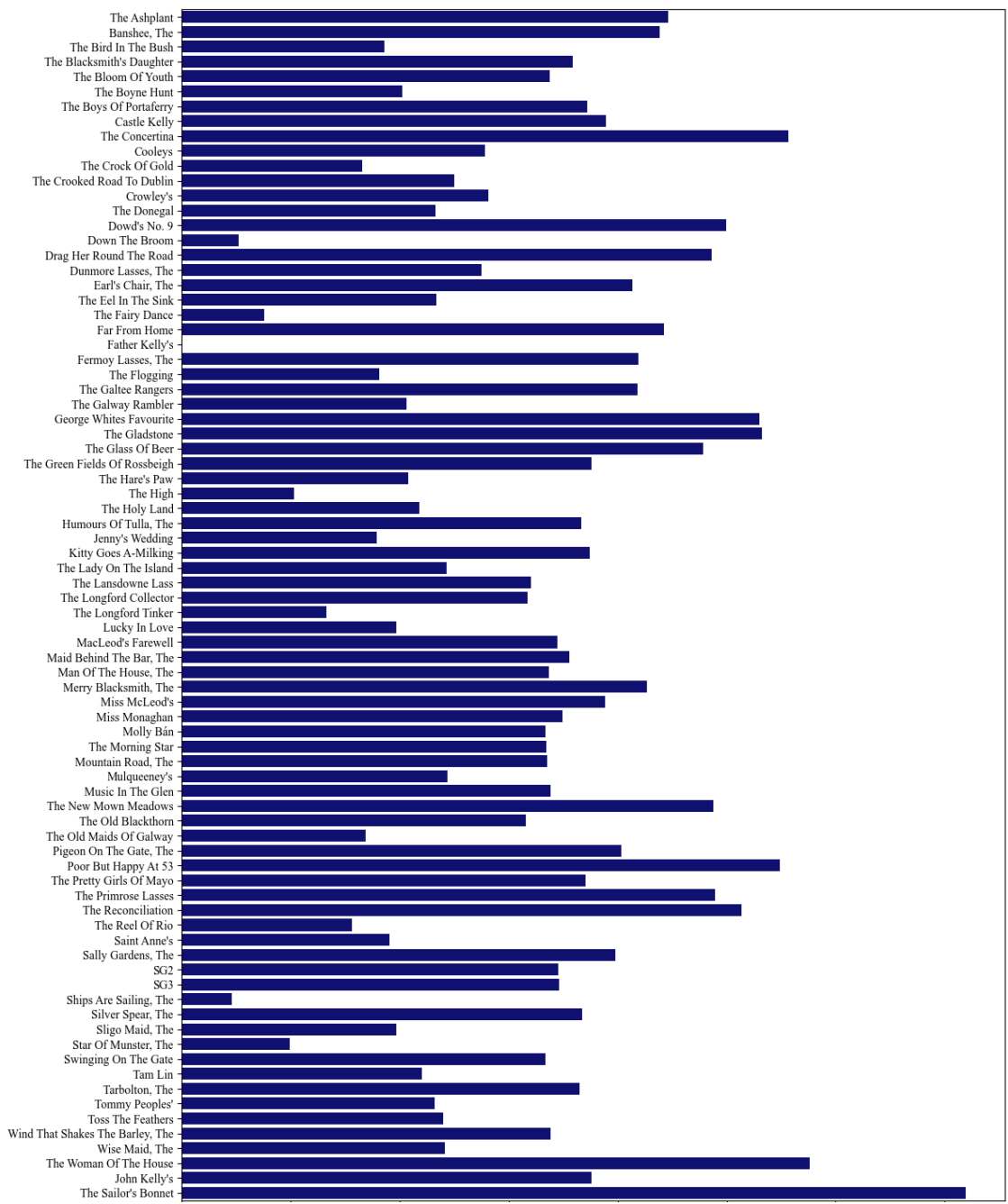

Figure 4.2: Gross Coherence.

Clearly, *The Sailor's Bonnet* was chosen by the model as the most likely match, proving in concept the fingerprinting capabilities of the wavelet transform. However, in pursuit of scientific inquiry, some additional factors will be discussed in the next sections to increase the model's decisiveness.

## 4.2 Prediction Accuracy

In implementing this kind of model for wavelet transform fingerprinting and tune identification, one comes across innumerable knobs and dials that can affect the results. To name a handful, we could fudge the sampling frequency, range of scales, image resolution, beats per minute (BPM), playtime duration, and harmonic overtones in synthesizers. These must be arbitrated to walk a delicate balance of multiple desirable properties for the model, including computational efficiency, predictive accuracy, ease of use, and explanatory power.

For example, convolving a thousand wavelets at scales ranging between 200 and 4000 Hz may have an unnecesarily high resolution in the frequency domain, dramatically increasing the compute time. Conversely, sending only fifty wavelets results in insufficient distinctions between the wavelengths. Most of the results in this work are from wavelet transforms with 200 scales, making 200 the maximum resolution in the frequency domain.

In Chapter 2, it was revealed that humans can hear pitches above 20,000 Hz, yet all of our results display axes capped at 4000 Hz. Furthermore, we mentioned that most Irish music is explicitly written below 1200 Hz, which raises the question of why this range was chosen. Why not cap it at 1200 to capture only the frequencies that are expressed symbolically? Or why not 20,000, and see a more complete profile of the sounds we are experiencing?

Here, we are aiming to balance prediction accuracy with explanatory power. Recording the tunes at an 8000 Hz sampling rate is approaching the lowest frequency where the tunes are still recognizable, and 4000 Hz is the Nyquist frequency at this rate. A cap of 4000 Hz allows us to see some of the most prevalent overtones of our recorded and synthetic waveforms, while still maximizing the frequencies that are necessary to compare.

If one were to cap the scaling at 1200 Hz, there would be a noticeable boost in the accuracy of the model. The coherence array and resulting summations are shown in Figures 4.3 and 4.4, respectively.

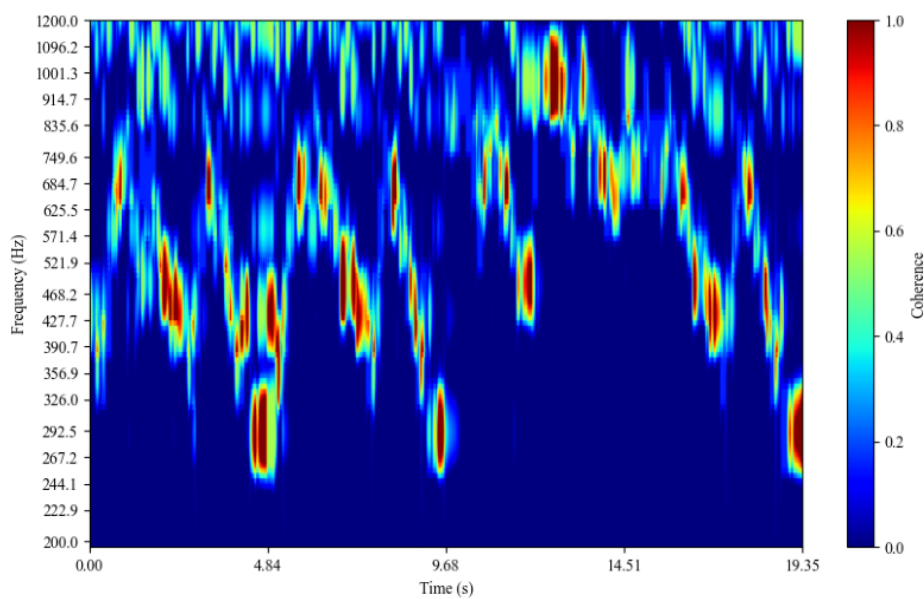

Figure 4.3: *The Sailor's Bonnet* Coherence in focused frequency range.

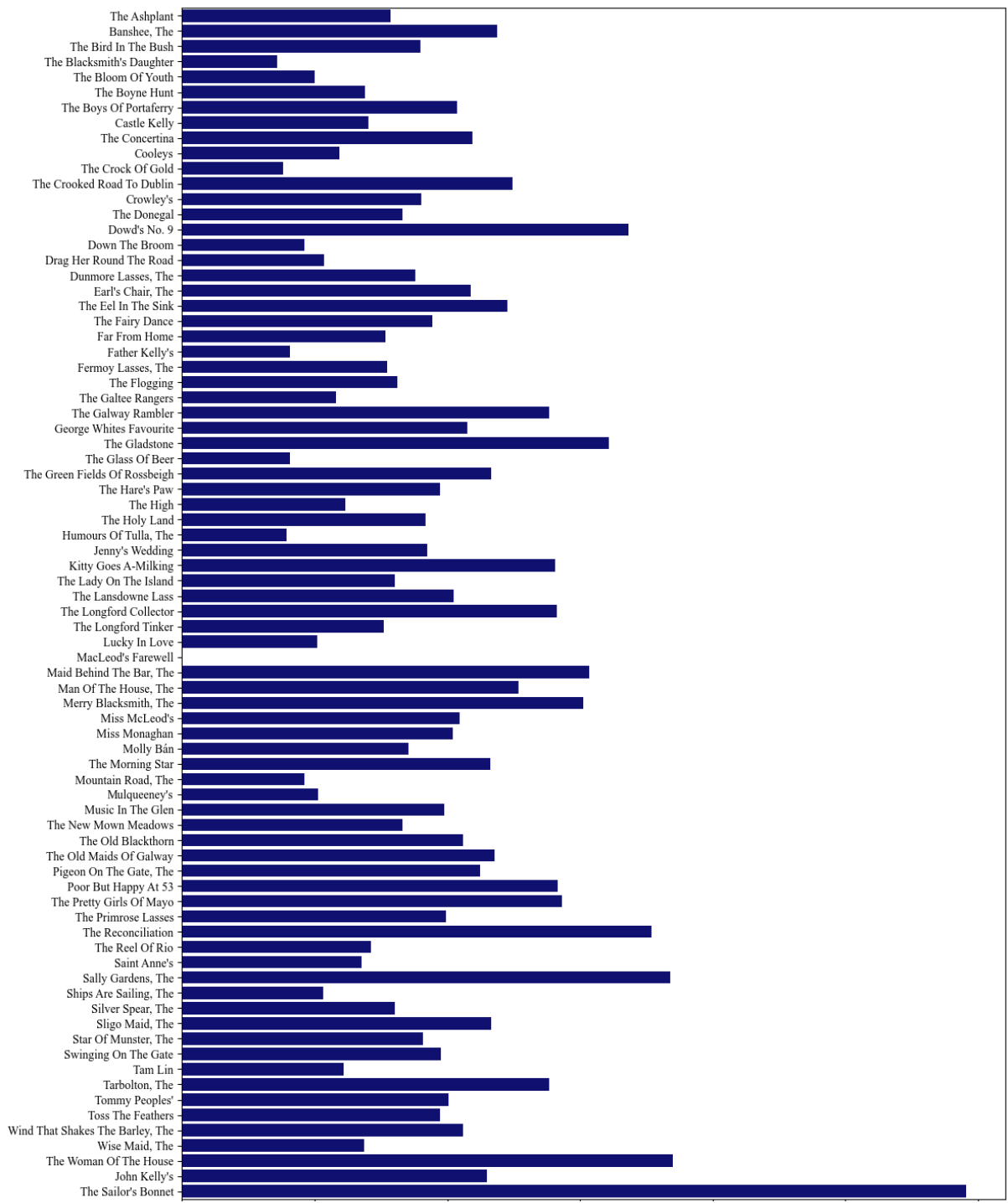

Figure 4.4: *The Sailor's Bonnet* resultant sums from low range comparisons.

As a quick footnote, the results in Figures 4.2, 4.4 and 4.10 have been calibrated to the tune with the lowest coherence. For uncalibrated results, see Appendix B.

## 4.3 Instrumentation

As the coherence transforms are scaled logarithmically, the details at the higher frequencies are lost when compared to the low end of the spectrum. Those high regions being where the harmonics live, it would be expected for the model to be more capable of comparisons between spectrograms that omit the overtones.

When calculating the coherence, each input has its own source of overtones. For the synthetic tune, the harmonics are created by the process shown in Figure 3.4. These additional frequencies are modeled after the timbre of a grand piano, whose details are described in Appendix A.

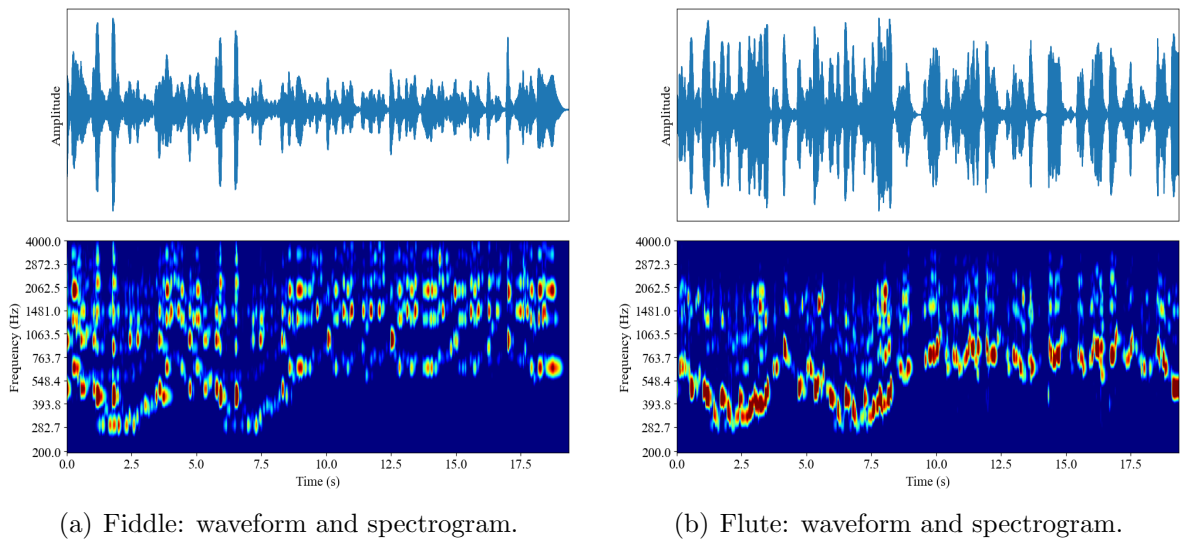

(a) Fiddle: waveform and spectrogram.  (b) Flute: waveform and spectrogram.

Figure 4.5: Wavelet transforms of the *Ships are Sailing* reel, recorded on fiddle and flute.

The live counterpart data will carry the natural harmonic frequencies of whichever instrument was being recorded. For example, observe the recording of *Ships are Sailing*, played on both the fiddle and the Irish flute[1], as seen in Figure 4.5. Clearly, the bowed string instrument has a more pronounced overtone profile than the flute, whose strongest frequencies lie solidly within the region corresponding to the written tune, which is shown in Figure 4.6.

From the waveforms alone in Figure 4.5, several differences between the instruments are immediately apparent. Notably, the flute waveform contains distinct moments of pure silence, corresponding to the flutist pausing to take a breath. In contrast, the fiddle produces a continuous noise, with some amplitude always present. Additionally, this suggests that the tune may not have been played exactly according to the score, which calls for a *rest* at the end of the A part, as indicated in Figure 4.6, Measure 8.

In the previous section, we discussed how adjusting the frequency scope of the coherence transform can affect the model's ability to identify tunes. Specifically, we observed that restricting the highest frequency to match the range expected from standard score notation may improve the decisiveness of the results. Based on this prior, we hypothesize that the flute may serve as a more effective instrument for tune identification, due to its strong emphasis on fundamental frequencies.

---

[1] Special thanks to Jeremy Jenkinson for providing the flute playing.

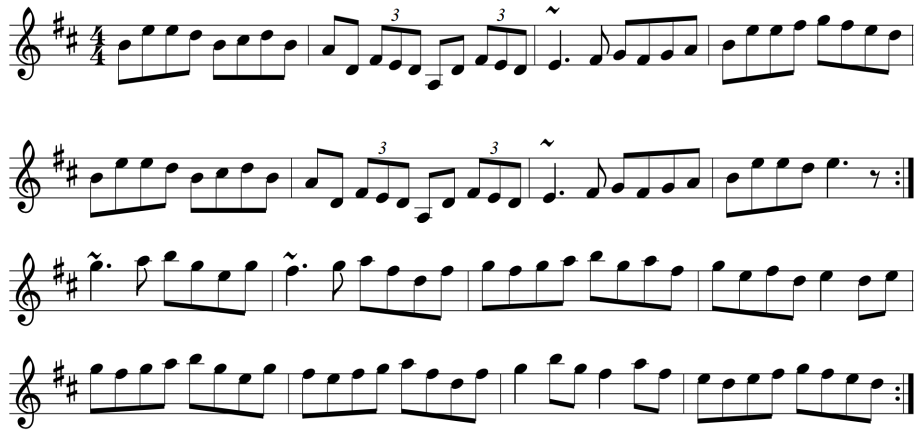

Figure 4.6: Musical Score for the *Ships are Sailing* reel.

To examine this hypothesis, consider Figure 4.7, which displays the summed coherence scores. The model appears to be more decisive for the flute recording than for the fiddle, as indicated by the significantly higher coherence score for *Ships Are Sailing* in the flute recording. Naturally, this is just a single comparison between two recordings, and does not constitute a rigorous conclusion about identifiability across instruments. However, it is worth noting that many players anecdotally report that apps such as TunePal tend to be more reliable in identifying tunes played on the flute or whistle.

Work by Bryan Duggan (creator of TunePal), for his *Machine Annotation of Traditional Tunes* (MATT2) algorithm makes use of an *onset detection* that accommodates for phrasing in the concert flute and tin whistle. This special treatment of the expressiveness in phrasing, along with ornamentation, long notes, and transposing by octave "results in a statistically significant improvement in annotation accuracy over approaches that do not accommodate expressiveness" [24].

Bellows instruments, such as the concertina and accordion, use reeds that provide a clean, fundamental-heavy tone. This should facilitate reliable identifiability in models such as ours, though these instruments use multi-reed systems, where each button corresponds to two (or sometimes three) reeds, slightly out of tune with each other to produce an oscillating, or slight vibrato effect. Further discussion of the timbres in Irish traditional instruments is provided in Appendix A.

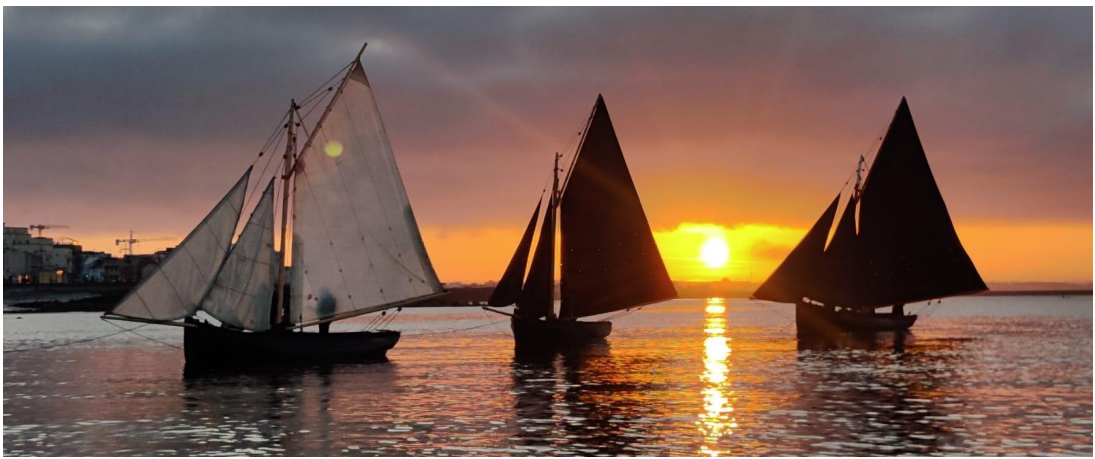

**Ships are Sailing at Sunrise**
Galway Hooker Sailing Club, 2025

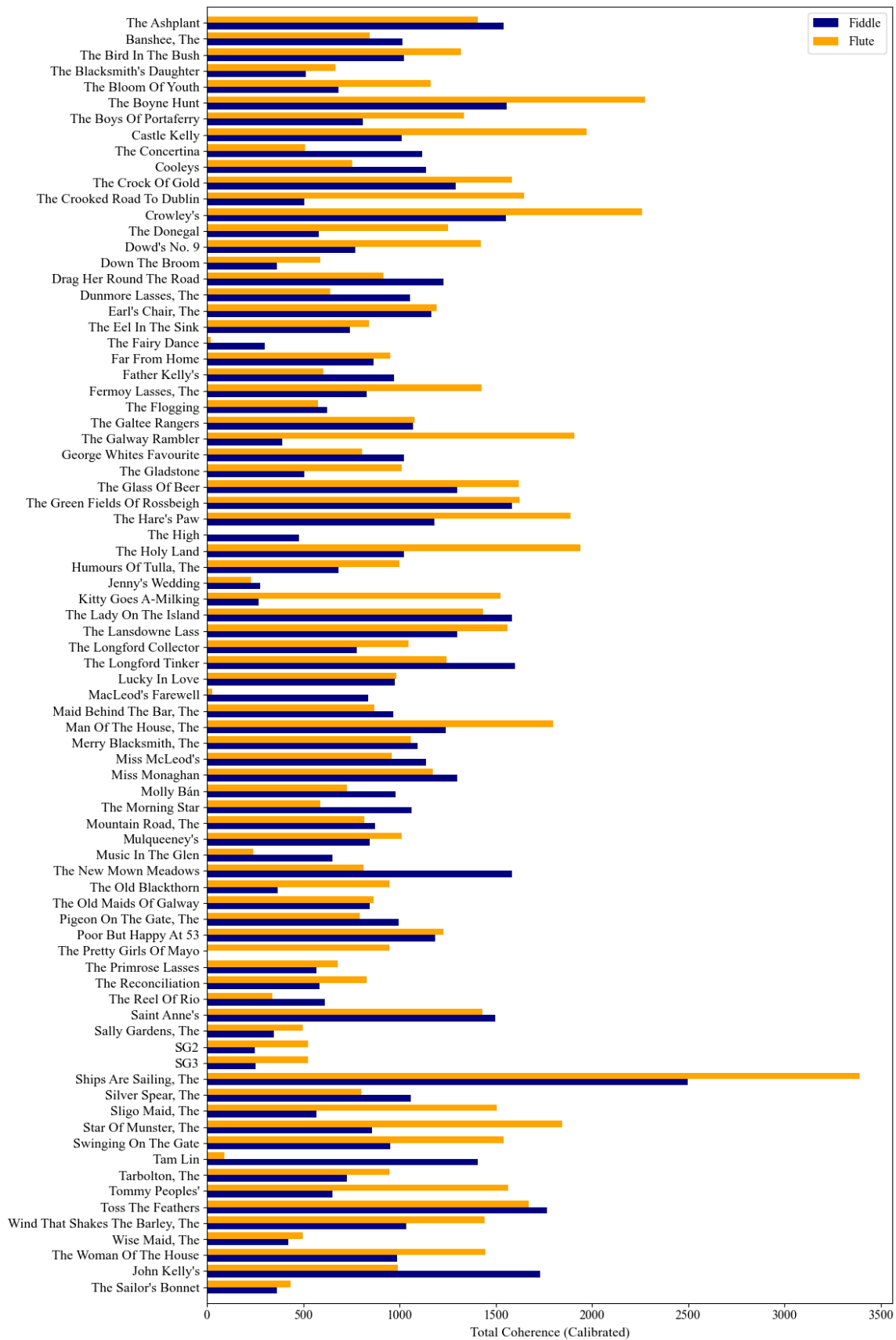

Figure 4.7: Musical Score for the *Ships are Sailing* reel.

## 4.4 Tune Biases

As this model's capabilities are explored, it can be observed that certain tunes are favored when it comes to the selection process. Although quite reliable for all tunes sampled, there are some recurring titles that show up as second or third choices, or are selected when the recorded input is not a recognizable tune. To probe these biases, two control data series are constructed. The first signal, $a = \{a_n\}_{n=1}^{N}$ where $a_n = 0$, is completely blank. The second, $b = \{b_n\}_{n=1}^{N}$ is randomly generated white noise at frequencies ranging from 200 to 3000 Hz for 19.2 seconds. Both $a$ and $b$ are plotted in Figure 4.8.

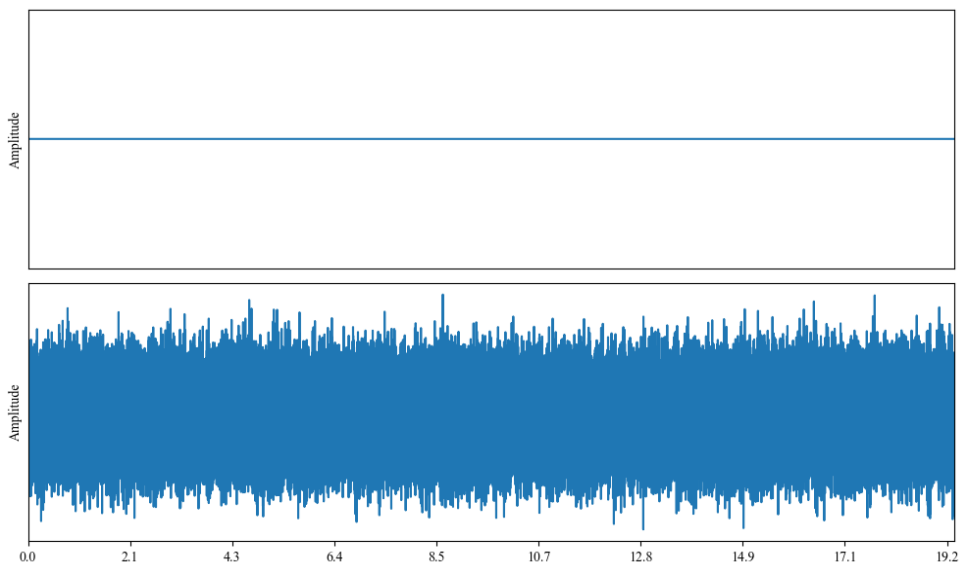

Figure 4.8: Controls $a$ (top) and $b$ (bottom).

The spectrograms produced by controls $a$ and $b$ are shown in Figure 4.9, and the coherence results are shown in Figure 4.10. These results synchronize with anecdotal observations, in which tunes such as *The Glass of Beer*, *The Concertina Reel,* or *The Mountain Road* tend to score highly, regardless of what is being recorded. However, these differences are amplified by the calibration, and the uncalibrated results in Appendix B show how small these biases are in reality.

These differences could arise from a number of factors, such as tune complexity or wavelength scaling. Because lower frequencies are emphasized by our logarithmic scaling of the y-axis, they will take up more space in our spectrograms and contribute more to the overall coherence. This means that tunes with more low pitches may be slightly advantaged.

Another explanation for the differences could be due to the smoothing operator that was implemented in the coherence calculation in Equation (3.4.1). Longer or repeated notes in a tune can blend together in the spectrograms, to create deeper red regions when the smoothing takes effect, benefiting tunes with more repetition. This could explain the apparent coincidence between biased tunes and tunes of low *Kolmogorov Complexity*, as catalouged by Mc Gettrick et al. [13].

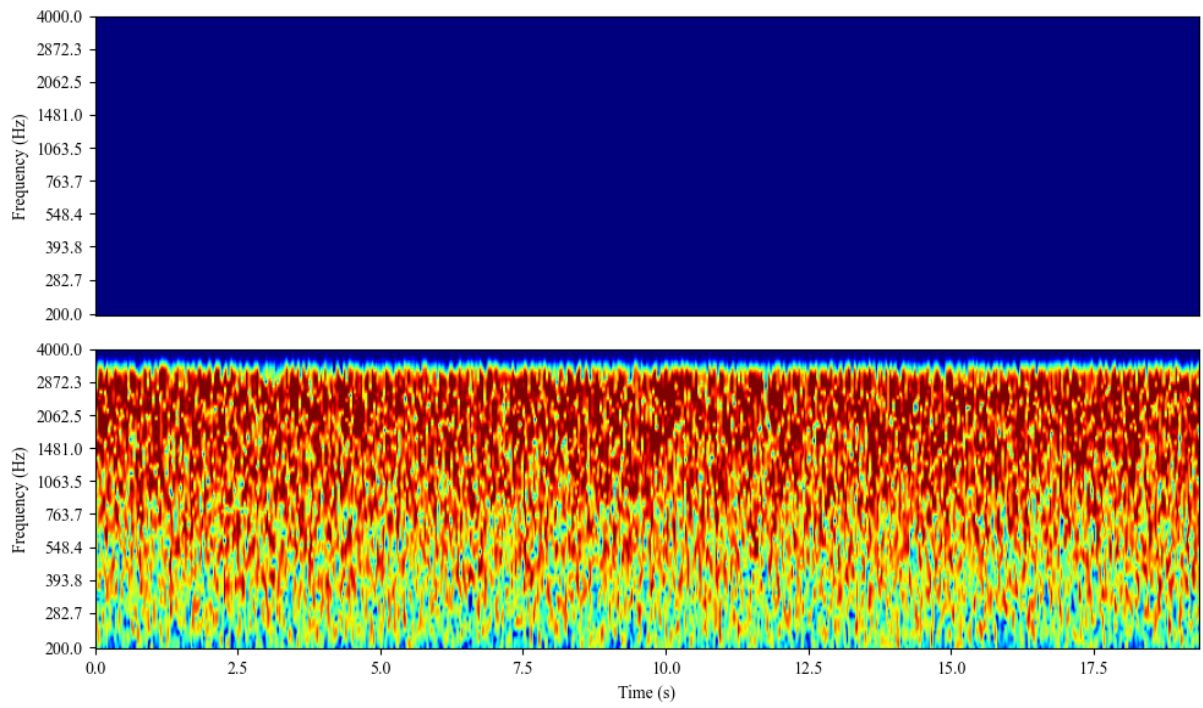

Figure 4.9: Spectrograms for $a$ (top) and $b$ (bottom)

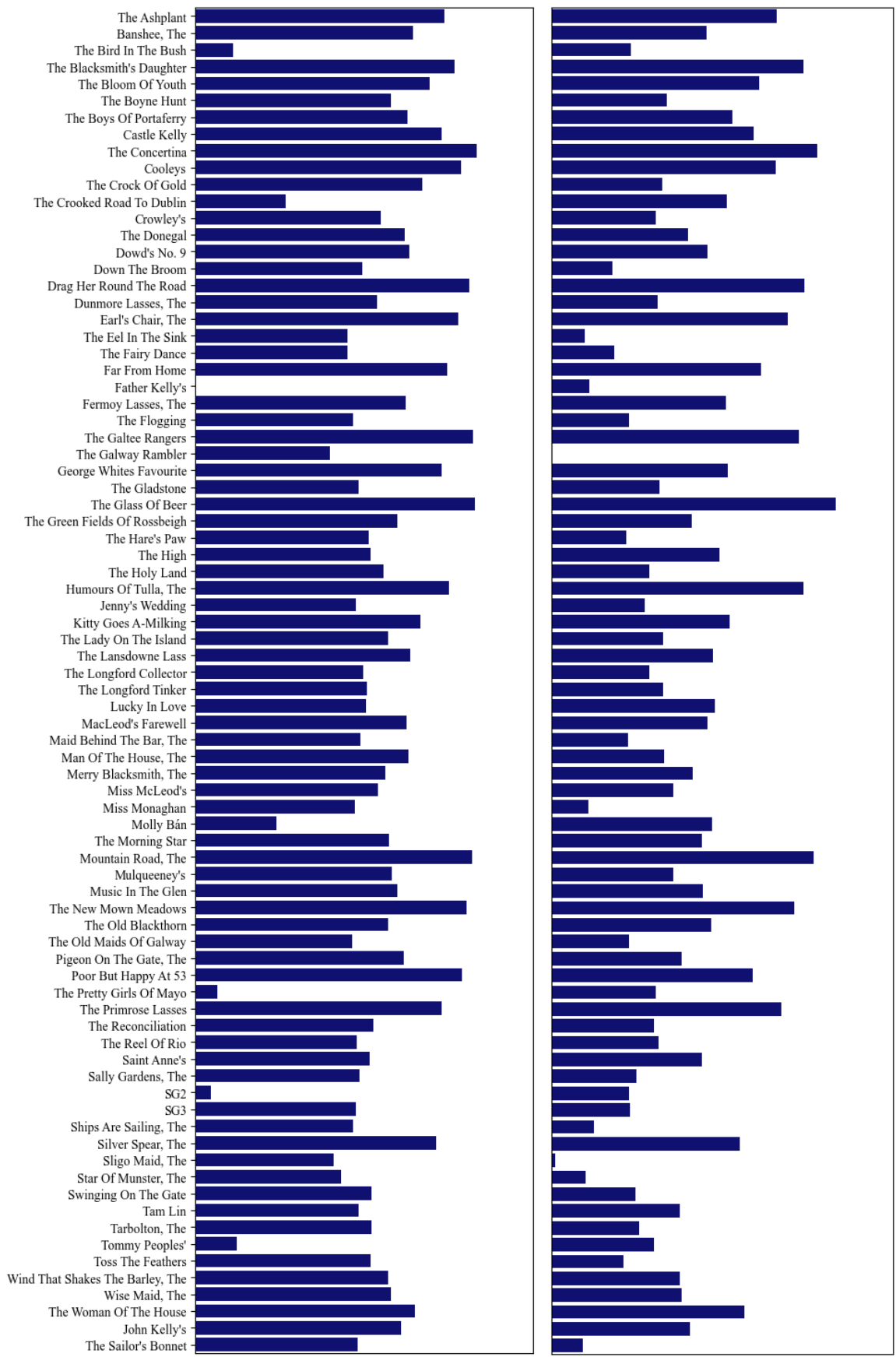

Figure 4.10: Total coherence for Controls $a$ (left) and $b$ (right).

# Chapter 5

# Discussion

> Additional details and applications are explored, along with some notes about algorithm efficiency and closing remarks.

## 5.1 Phase

Although the previous example demonstrates the accuracy of the wavelet transform model, it does not fully exploit the capabilities of wavelet coherence analysis. For the continuous wavelet transform (CWT), the Morlet wavelet is most commonly used because it produces both real and imaginary components in the wavelet coefficients. This allows us to determine not only which frequencies are present in the signals, but also their relative phase alignment over time.

To illustrate this, consider the example of sine and cosine waves oscillating at 25 Hz in Figure 5.1. These basis functions are exactly 90 degrees out of phase, a relationship that should be reflected when applying the *cross-wavelet transform* (XWT), as defined in Equation (3.4.2). The resulting quiver plot, overlaid on the coherence array in Figure 5.2, visualizes this phase relationship.[1]

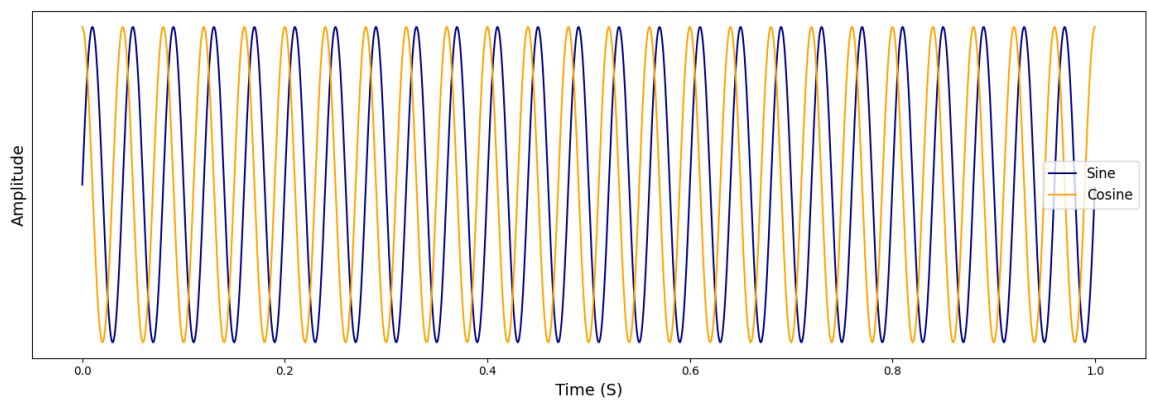

Figure 5.1: Sine and cosine functions at 25 Hz.

---

[1] At these low frequencies, the *cone of influence* (COI) noticeably affects the boundaries. More information on the COI can be found in Appendix B.

Because the cosine function's wavelet coefficients are passed as the second argument to the XWT, they are complex-conjugated, while the sine function's coefficients remain unchanged. This results in arrows pointing downward, indicating that the sine wave lags behind the cosine. Reversing the order of the arguments would cause the arrows to point upward instead.

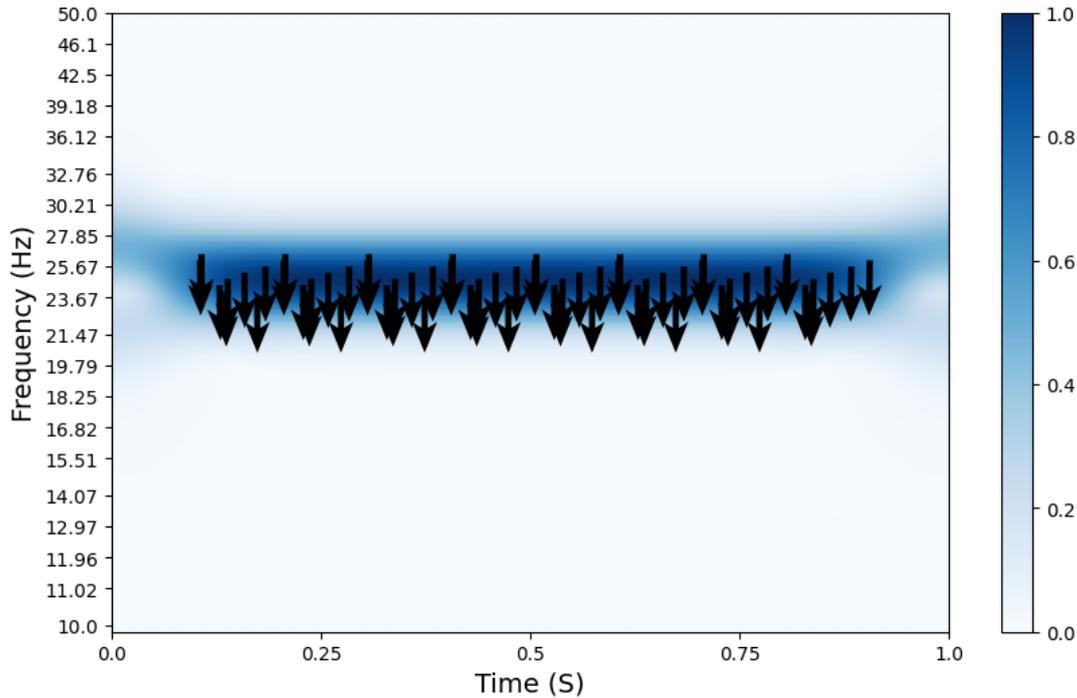

Figure 5.2: XWT quiver plot over coherence between sine and cosine functions.

Recall that the Fourier transform encodes phase alignment as the argument of its complex-valued coefficients. This can be a significant advantage in analyzing signals where phase relationships are important but not readily apparent in the time domain. However, because the Fourier transform is a global operator, a phase shift of $\theta$ radians may appear only as a rotation between complex coefficients. This manifests as two frequency components at an angular difference of $\theta$ radians in the complex plane, telling us that multiple phases are present but providing no information about when—or if—a shift occurs.

The beauty of wavelet coherence lies in its ability to properly define these shifts. By localising phase information in both time and scale, wavelet coherence can reveal exactly when a phase shift occurs. This is crucial for determining causal relationships from the frequency spread, and we will see applications of this in the next section.

## 5.2 Alternative Applications of the Coherence Model

The phase arrays contain incredibly valuable information when we are interested in the relationship between our cycling elements. To pull an example from the playground of econometrics, imagine we have a hypothesis that the markets of two commodities are related somehow, and the price of the first has an effect on the price of the second. Wavelet coherence is perfectly suited for this task because it tells us:

- When these commodities are tracking each other,
- At what frequencies the prices are cycling,
- The lag period between each cycle.

The New York Stock Exchange (NYSE) contains thousands of assets with prices being updated every second. As an example, we take the time series data from Colgate-Palmolive Co (**CL**) and Eversource Energy (**ES**), between June 7, 2024 and June 7, 2025, sampling every day the stock market is open, totaling 250 days. These are the same two tickers selected in by N. Picini [14] to show how oil futures follow the overall market. The coherence between these two time series can be seen in Figure 5.3.

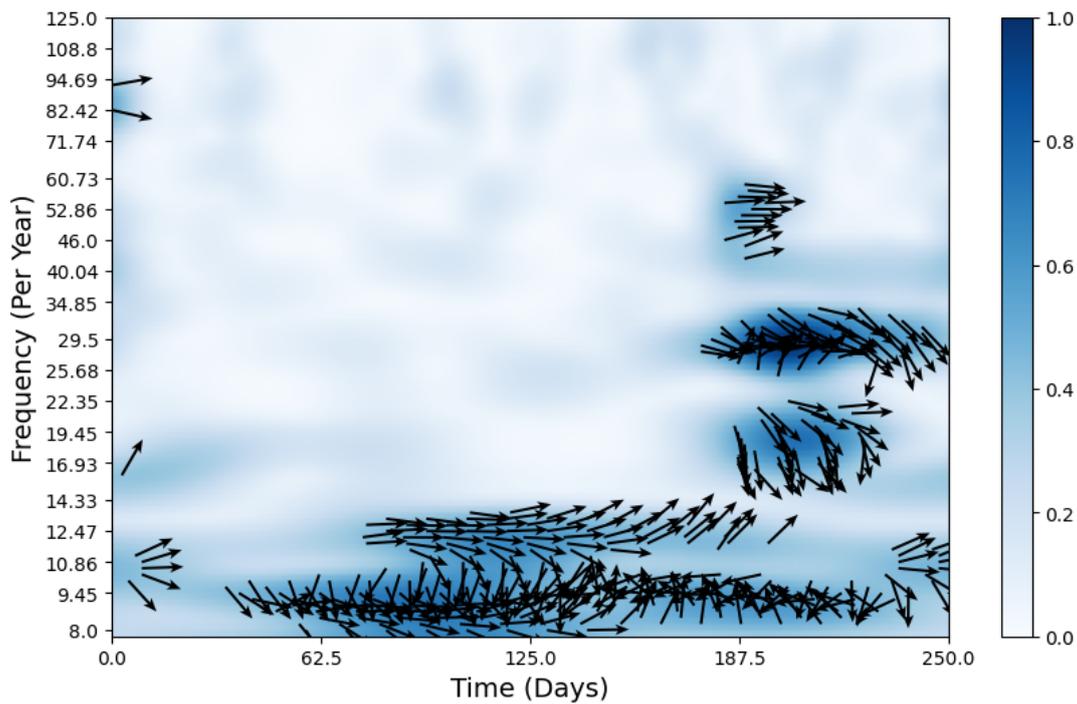

Figure 5.3: NYSE tickers **ES** and **CL** coherence with phases, June 2024 to June 2025.

Another domain with growing interest in wavelet coherence methodology is electroencephalography (EEG). Coherence is widely used to study the dependencies between electrodes measuring brain activity at specific regions, as seen in work by Ieracitano et al. [16]. To demonstrate an example with EEG data[2], we pull 60 seconds of brain activity of a research subject at the frontal midline (**Fz**) and parietal midline (**Pz**) electrodes. The EEG data are displayed in Figure 5.4, and coherence results are displayed in Figure 5.5.

---

[2] Data sourced from Subject 2, condition 253 of the *semantic task with 5-subjects* study on EEGLAB [15].

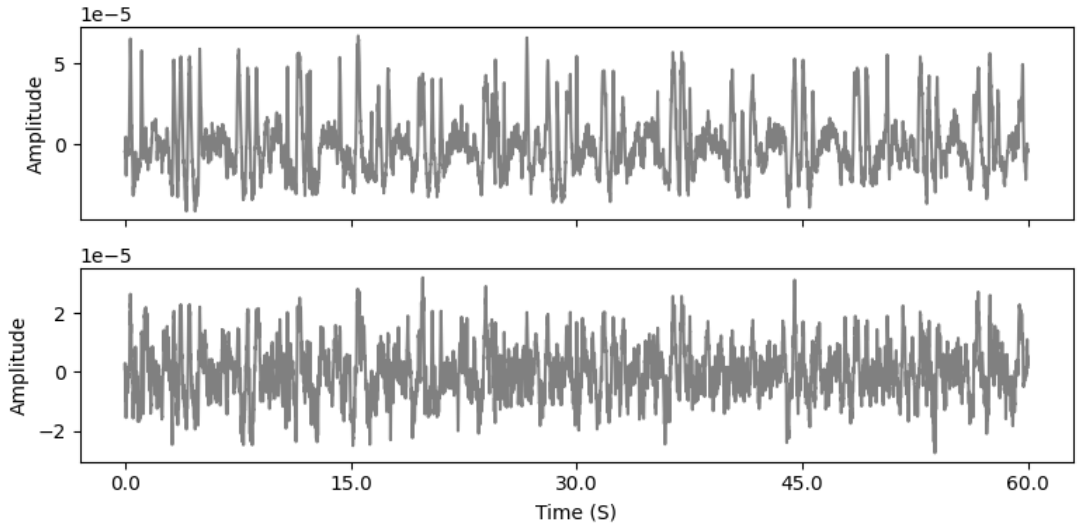

Figure 5.4: **Fz** (top) and **Pz** (bottom) electrodes measuring brain activity.

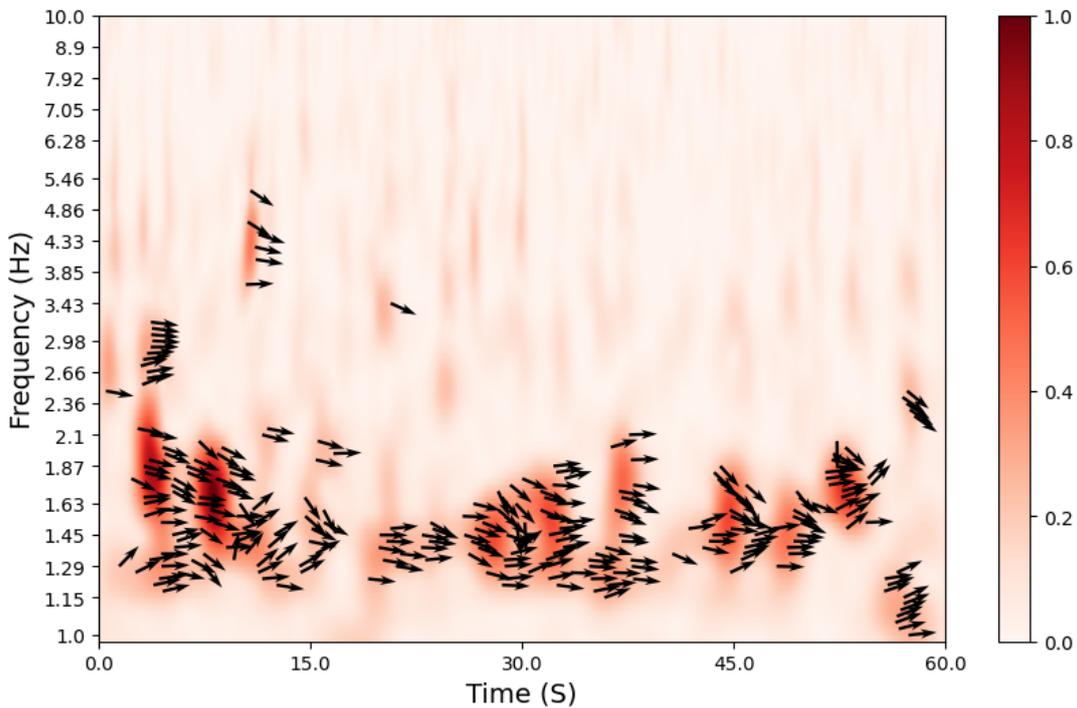

Figure 5.5: Coherence between brain activity at **Pz** and **Fz** electrodes.

What the results in Figures 5.3 and 5.5 imply specifically is contained within their respective scientific domains, and thus meaningful interpretation remains outside the scope of this work. Our motivation for applying the coherence model to these datasets stems from the pre-existence of the aforementioned literature itself, in which coherence is applied to such domains.

## 5.3 Computation

The wavelet transform of Section 2.5 is written entirely in the time domain. While this is helpful for understanding how we arrive at our finalized time-frequency products, the direct convolution is severely impractical. Equation (2.5.1) is often called the *naive* implementation of CWT. The convolution of the discrete signal $x(t)$ and a discretized and time-localized wavelet function $\psi(t)$ (both of length $N$) requires $N^2$ computations, so the compute time scales quadratically. Because we must produce this convolution at each scale, the computational complexity of the naive implementation is written as:

$$\mathcal{O}(S \cdot N^2) \tag{5.3.1}$$

Where $S$ is the number of scales.

The naive implementation is easily improved upon by converting both the signal and the wavelet into the frequency domain via FFT, then the two vectors can simply be multiplied and sent back into the time domain by iFFT. To explore the savings in computational complexity by this method, we follow these steps:

1. **Compute the FFT of the signal** $x(t)$:

$$\hat{x} = \text{FFT}(x) \quad \text{Cost: } \mathcal{O}(N \log N) \tag{5.3.2}$$

2. **For each scale** $s_i$ $(i = 1, \ldots, S)$:

    (a) Generate the wavelet $\psi_{s_i}(t)$ at scale $s_i$, Cost: $\mathcal{O}(N)$

    (b) Compute the FFT of the scaled wavelet:

$$\hat{\psi}_{s_i} = \text{FFT}(\psi_{s_i}) \quad \text{Cost: } \mathcal{O}(N \log N) \tag{5.3.3}$$

    (c) Perform pointwise multiplication in the frequency domain:

$$\hat{w}_{s_i} = \hat{x} \cdot \hat{\psi}_{s_i} \quad \text{Cost: } \mathcal{O}(N) \tag{5.3.4}$$

    **Note:** $\hat{\psi}_{s_i}$ and $\hat{x}$ contain complex numbers.

    (d) Compute the inverse FFT to obtain the convolution result:

$$w_{s_i}(t) = \text{iFFT}(\hat{w}_{s_i}) \quad \text{Cost: } \mathcal{O}(N \log N) \tag{5.3.5}$$

3. **Total cost over all scales**:

$$\mathcal{O}(N \log N) \text{ (for signal FFT)} +$$
$$S \cdot [\mathcal{O}(N) + \mathcal{O}(N \log N) + \mathcal{O}(N) + \mathcal{O}(N \log N)] = \mathcal{O}(S \cdot N \log N) \tag{5.3.6}$$

Additionally, the results of step 2-b can be easily precomputed and cached or approximated within the frequency domain without performing the full FFT at every scale. Therefore, in any practical implementation of the FFT wavelet transform, virtually all of the compute time is consumed by step 2-d, where we take the inverse Fourier transform at each scale. The next section will discuss some recent findings that circumvent this last step and accelerate the performance of the model used in this work.

## 5.4 Performance

In 2022, researchers P.A. Arts and L. van den Broek published the *Fast Continuous Wavelet Transform* (fCWT) algorithm, with dramatically positive ramifications for wavelet transform computation. The fCWT enables the coherence model used in this work by scaling down the amount of compute time required for long signals, focusing on the inverse Fourier transform step in the FFT-based implementation of the wavelet transform.

To give an overly brief technical summary of what is going on in the background, this efficient algorithm caches precomputed look-up tables of the iFFT and can apply strategic downsampling to the time-frequency complex matrix prior to the scale dependent operations, allowing large-scale reductions in overall computational complexity. A comprehensive explanation of the fCWT algorithm can be found in its original publication in *Nature Computational Science* [10].

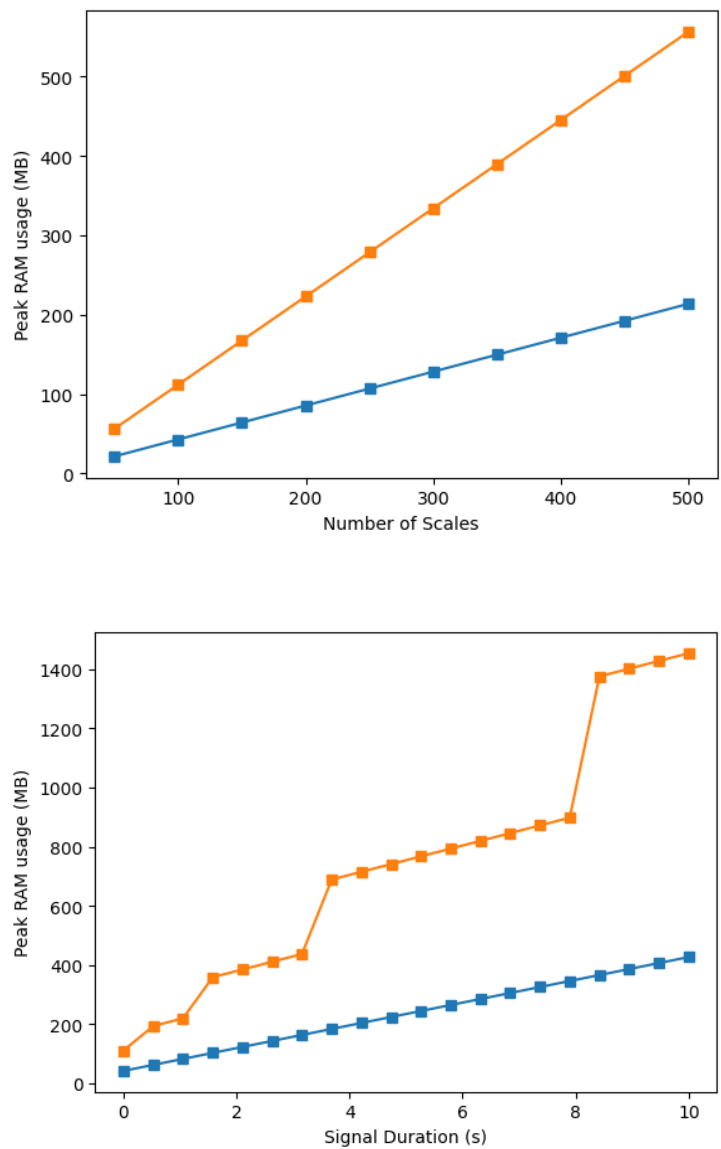

Figure 5.6: Efficiency comparison in memory usage. Top: increasing frequency resolution, bottom: increasing number of samples. Legend: ■ - Our model, ■ - PyCWT benchmark.

To illustrate how much we can save by these techniques, we construct two signals that are just pure noise at 8000 Hz SR, both similar to Signal $b$ in section 4.4, and produce the coherence matrix via our model, in parallel with the PyCWT library's wavelet coherence transform (WCT) method [22], which serves as a benchmark. We are interested in how the efficiency is affected in both of our domains of interest, so we run two experiments: Figure 5.4 shows how increasing resolution in the frequency and time domains affects the memory usage.

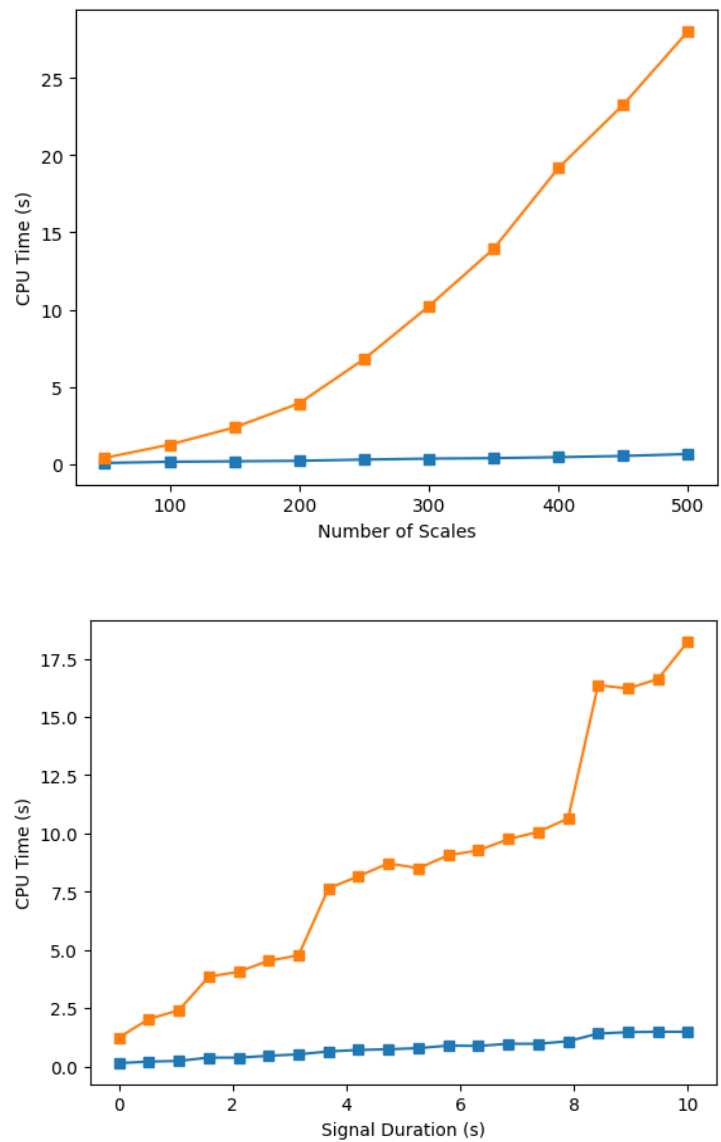

Figure 5.7: Efficiency comparison in process time. Top: increasing frequency resolution, bottom: increasing number of samples. Legend: ■ - Our model, ■ - PyCWT benchmark.

Figure 5.7 shows the same two experiments as above, only our metric is now processing time instead of memory usage. These results closely replicate the performance findings in the original fCWT publication, showing the large portion of coherence computation being consumed by the supporting wavelet transforms. Additionally, the step-wise behavior observed when increasing signal duration is a well documented phenomena in FFT based implementations, which are most efficient at transforming signals whose input lengths are $2^n, n \in \mathbb{N}$. As a result, such algorithms will pad signals to the next power of 2 [21].

## 5.5 Conclusion

This work set out to explore wavelet-based time-frequency decompositions to serve as identifying fingerprints for traditional Irish music, for both live and symbolic inputs. By transforming each data type into frequency spectra for analysis, we demonstrated that wavelet coherence can be used to accurately identify recorded tunes by matching them against a library of synthetically generated waveforms derived from ABC notation. This approach was found to be efficient and reliable in a variety of instruments, although many opportunities for model optimization remain.

Beyond music identification, this research demonstrates the flexibility of the coherence model with clear applications in other domains. In EEG signal processing, wavelet coherence between electrodes reveals patterns of brain connectivity across different regions. In financial price fluctuation data, coherence can illuminate historic market co-movements and lead-lag relationships between assets over time. These diverse applications highlight the generality of the wavelet-based approach and suggest that the methodology outlined here may be useful for other tasks in non-stationary environments.

Looking forward, we aim to apply the methods explored here as modules in larger frameworks. This will involve interpreting the time series wavelet outputs as extracted features to feed into artificial neural networks, or applying machine learning over the coherence arrays to determine when coherence is meaningful and when it is random. This mode of application for coherence is hypothesized to prove useful for anomaly detection as well, where disparate coherent regions in the time-frequency space between null, unrelated events can be flagged for further investigation.

# Appendix A

# Code

Python printouts of several key functions for the tune identification model.

## A.1 Coherence

```python
import numpy as np
import fcwt
from scipy.ndimage import gaussian_filter

def smooth(data, sigma=(2, 2), mode='nearest'):
    return gaussian_filter(data, sigma=sigma, mode=mode)

def wavelet_coherence(signal1, signal2, highest, lowest, nfreqs,
   frame_rate, window=(2, 2)):
    freqs, coeffs1 = fcwt.cwt(signal1, frame_rate, lowest,
        highest, nfreqs, nthreads=4)
    freqs, coeffs2 = fcwt.cwt(signal2, frame_rate, lowest,
        highest, nfreqs, nthreads=4)

    # Squared modulus of wavelet coefficients
    S1 = np.abs(coeffs1) ** 2
    S2 = np.abs(coeffs2) ** 2
    xwt = coeffs1 * np.conj(coeffs2)

    numerator = np.abs(smooth(S12)) ** 2
    denominator = smooth(S1) * smooth(S2)

    # Prevent division by zero
    coherence = np.zeros_like(numerator)
    nonzero = denominator > 0
    coherence[nonzero] = numerator[nonzero] /
        denominator[nonzero]

    return coherence, freqs, xwt
```

Listing A.1: Wavelet Coherence Computation.

To compute the wavelet transforms seen in this work, we can use the seamlessly integrated **fcwt** python implementation, as shown in Listing A.1. This import runs all computations in C++ on the back-end. This function returns the coherence array, a vector of frequencies, and the *cross-wavelet transform*, saved as the object **xwt**, as mentioned in Section 5.1.

## A.2  Musical Note Generators

Given a frequency and a sampling rate, we can use the following function to generate a sine wave to match our desired duration of the note. If the envelope is requested, this can be controlled with the boolean ***envel***. More information on envelopes will be found in the next section.

```python
def sine_note(frequency, duration, sample_rate, envel = True):
    t = numpy.linspace(0, duration, int(sample_rate * duration),
        endpoint=False)

    # Fundamental frequency
    wave = numpy.sin(2 * numpy.pi * frequency * t)
    adsr = [0.01, 0.2, 0.2, 0.7]
    adsr = [adsr[0]]+[(duration-adsr[0])*(x/sum(adsr[1:])) for x
        in adsr[1:]]
    # ADSR envelope (if desired)
    envelope = adsr_envelope(adsr[0],adsr[1],adsr[2],adsr[3],.7,
        sample_rate, duration)
    if envel:
        wave *= envelope

    # Normalize wave
    wave = wave / numpy.max(numpy.abs(wave))

    return wave
```

Listing A.2: Sine note generator function

To better simulate the sound of a real instrument, we can add the second, third, and fourth harmonics—along with some non-linearity—based on the harmonic profile of a piano.[1]

---

[1]From Weinreich [18]: "Measured Steinway D-274 spectra show the $2^{\text{nd}}$ harmonic at –4 dB ($\approx$ 0.6×), the $3^{\text{rd}}$ at –8 dB ($\approx$ 0.4×), and the $4^{\text{th}}$ at –14 dB ($\approx$ 0.2×) relative to the fundamental."

```python
def piano_note(frequency, duration, sample_rate):
    t = numpy.linspace(0, duration, int(sample_rate * duration),
        endpoint=False)

    # Fundamental frequency
    wave = numpy.sin(2 * numpy.pi * frequency * t)

    # Harmonics
    wave += 0.6 * numpy.sin(2 * numpy.pi * 2 * frequency * t)  #
        Second harmonic

    wave += 0.4 * numpy.sin(2 * numpy.pi * 3 * frequency * t)  #
        Third harmonic

    wave += 0.2 * numpy.sin(2 * numpy.pi * 4 * frequency * t)  #
        Fourth harmonic

    # Harmonic non-linearity
    wave += 0.05 * (wave**3)

    #  ADSR envelope
    adsr = [0.01, 0.2, 0.2, 0.7]
    adsr = [adsr[0]]+[(duration-adsr[0])*(x/sum(adsr[1:])) for x
        in adsr[1:]]

    envelope = adsr_envelope(adsr[0],adsr[1],adsr[2],adsr[3],.7,
        sample_rate, duration)
    wave *= envelope

    # Slight noise
    wave += 0.001 * numpy.random.normal(-1, 1, len(wave))

    # Normalize wave
    wave = wave / numpy.max(numpy.abs(wave))

    return wave
```

Listing A.3: Piano note generator function.

We can also simulate other instruments that are more popular in Irish music than the piano. The code below shows the generator for the tenor banjo, which contains overtones all the way through the sixth harmonic, a slightly detuned addition to the wave, and a much shorter attack in the ADSR envelope. Lines 8-12 specifically dictate how much each harmonic frequency contributes to the waveform.

```python
def banjo_note(frequency, duration, sample_rate):
    t = numpy.linspace(0, duration, int(sample_rate * duration),
        endpoint=False)

    # Fundamental frequency
    wave = numpy.sin(2 * numpy.pi * frequency * t)

    # Add harmonics
    wave += 0.7 * numpy.sin(2 * numpy.pi * 2 * frequency * t)  #
        Second harmonic
    wave += 0.5 * numpy.sin(2 * numpy.pi * 3 * frequency * t)  #
        Third harmonic
    wave += 0.3 * numpy.sin(2 * numpy.pi * 4 * frequency * t)  #
        Fourth harmonic
    wave += 0.2 * numpy.sin(2 * numpy.pi * 5 * frequency * t)  #
        Fifth harmonic
    wave += 0.1 * numpy.sin(2 * numpy.pi * 6 * frequency * t)  #
        Sixth harmonic

    # Slight detuning for resonance
    wave += 0.1 * numpy.sin(2 * numpy.pi * 1.01 * frequency * t)
    wave += 0.08 * numpy.sin(2 * numpy.pi * 0.99 * frequency * t)

    # Nonlinearity for brightness
    wave += 0.2 * (wave**3)

    # Banjo ADSR envelope
    adsr = [0.003, 0.08, 0.02, 0.1]
    adsr = [adsr[0]] + [(duration - adsr[0]) * (x /
        sum(adsr[1:])) for x in adsr[1:]]
    envelope = adsr_envelope(adsr[0], adsr[1], adsr[2], adsr[3],
        0.2, sample_rate, duration)

    wave *= envelope

    # Normalize wave
    wave = wave / numpy.max(numpy.abs(wave))
    return wave
```

Listing A.4: Banjo note generator function

## A.3 Envelopes

The following Python function generates an ADSR (attack, decay, sustain, release) envelope for shaping audio signals. To call this, and all other functions in this appendix, the ***numpy*** library must be imported.

```python
def adsr_envelope(attack, decay, sustain, release,
    sustain_level, sample_rate, duration):

    total_samples = max(1, int(sample_rate * duration))

    attack_samples = max(1, int(sample_rate * attack))

    decay_samples = max(1, int(sample_rate * decay))

    release_samples = max(1, int(sample_rate * release))

    sustain_samples = max(1, total_samples - attack_samples -
        decay_samples - release_samples)

    # Envelope phases
    attack_curve = numpy.linspace(0, 1, attack_samples)

    decay_curve = numpy.linspace(1, sustain_level, decay_samples)

    sustain_curve = numpy.ones(sustain_samples) * sustain_level

    release_curve = numpy.linspace(sustain_level, 0,
        release_samples)

    # Concatenate phases
    envelope = numpy.concatenate([attack_curve, decay_curve,
        sustain_curve, release_curve])

    return envelope[:total_samples]
```

Listing A.5: ADSR envelope function

## A.4  The Session Database

The following code shows how we use SQL queries within a Python environment to pull tunes from *thesession.org*. The database file contains a table for tune information linked to a separate table for aliases. With the knowledge that popular tunes often go by many names, we want to be able to pull tunes by relevant aliases listed on The Session.

```python
import sqlite3
import pandas as pd

conn = sqlite3.connect('thesession.db')

tunes = pd.read_sql_query("SELECT * FROM tunes", conn)

aliases = pd.read_sql_query("SELECT * FROM aliases", conn)

conn.close()

# will preprocess string before feeding it into initializer
def initializer(name,setting):
    # should return lowest setting, can request higher settings
        later
    matched = tunes[tunes['name'].apply(preprocess_string) ==
        name]
    if matched.empty == False:
        matched = matched.set_index('setting_id')
        matched.index = range(1,len(matched)+1)
        return matched.loc[setting], len(matched)
    else:
        id =
            int(aliases[aliases['alias'].apply(preprocess_string)
            == name]['tune_id'].iloc[0])
        matched = tunes[tunes['tune_id'] == str(id)]
        matched.index = range(1,len(matched)+1)

        return matched.loc[setting], len(matched)

def get_abc(name,setting):
    matched = tunes[tunes['name'] == name]
    matched = matched.set_index('setting_id')
    matched.index = range(1,len(matched)+1)

    return [matched['abc'][setting],matched['mode'][setting]]
```

Listing A.6: Code to pull tune data and metadata from The Session's Database.

# Appendix B

# Additional Products

*Relevant information that does not fit in the principal opus.*

## B.1 Cone of Influence

As the wavelet convolves near the boundaries of the signal, it becomes necessary to compute values at timesteps where the wavelet is partially outside the available time range. To maintain a consistent number of timesteps at each scale—and thus produce a rectangular resolution in the wavelet transform—we must pad the edges of the signal with zeros. This ensures that even the lowest-frequency wavelets, which are the widest in time, have the same number of valid steps as the highest-frequency wavelets.

At the low frequencies, this results in some "junk" frequencies being detected, as the wavelet is partially over the signal and partially over the padding. The regions in the wavelet transform where this takes effect are said to be outside of the *cone of influence*. To see how the cone of influence affects our interpretation of the coherence results, take a look at the stationary Signals A and B in Figure B.1. Signal A is the summation of a sine wave at 30 Hz and a cosine wave at 75 Hz, while Signal B is the inverse of A, with the sine contribution at 75 Hz and the cosine at 30 Hz.

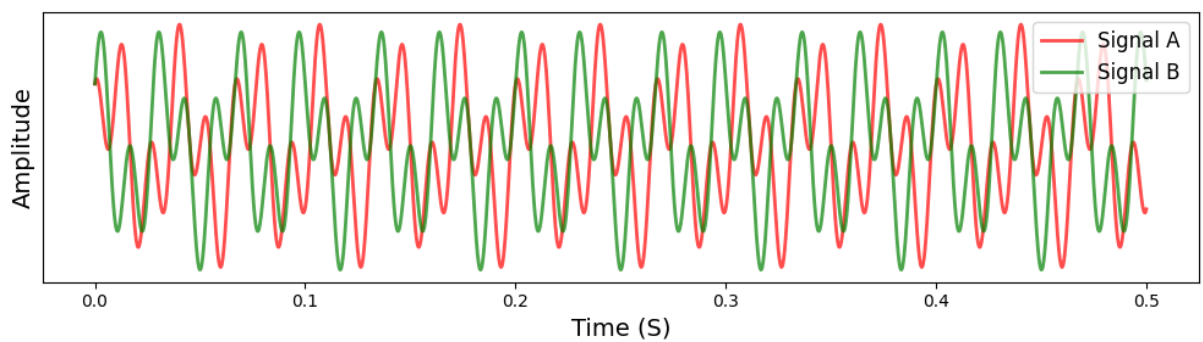

Figure B.1: Signals A and B.

For a quick description, this should result in two stripes across our spectrograms at the corresponding frequencies. At 75 Hz, the phase should be $-\pi/2$ radians, as the cosine of Signal A leads the sine of Signal B. Conversely, the 30 Hz stripe should reflect a phase of $\pi/2$ radians. This is what is shown in Figure B.2. The COI has a much heavier effect on the 30 Hz stripe as it is produced by a bulkier wavelet, causing the extent of the phase arrows to shrink in this region.

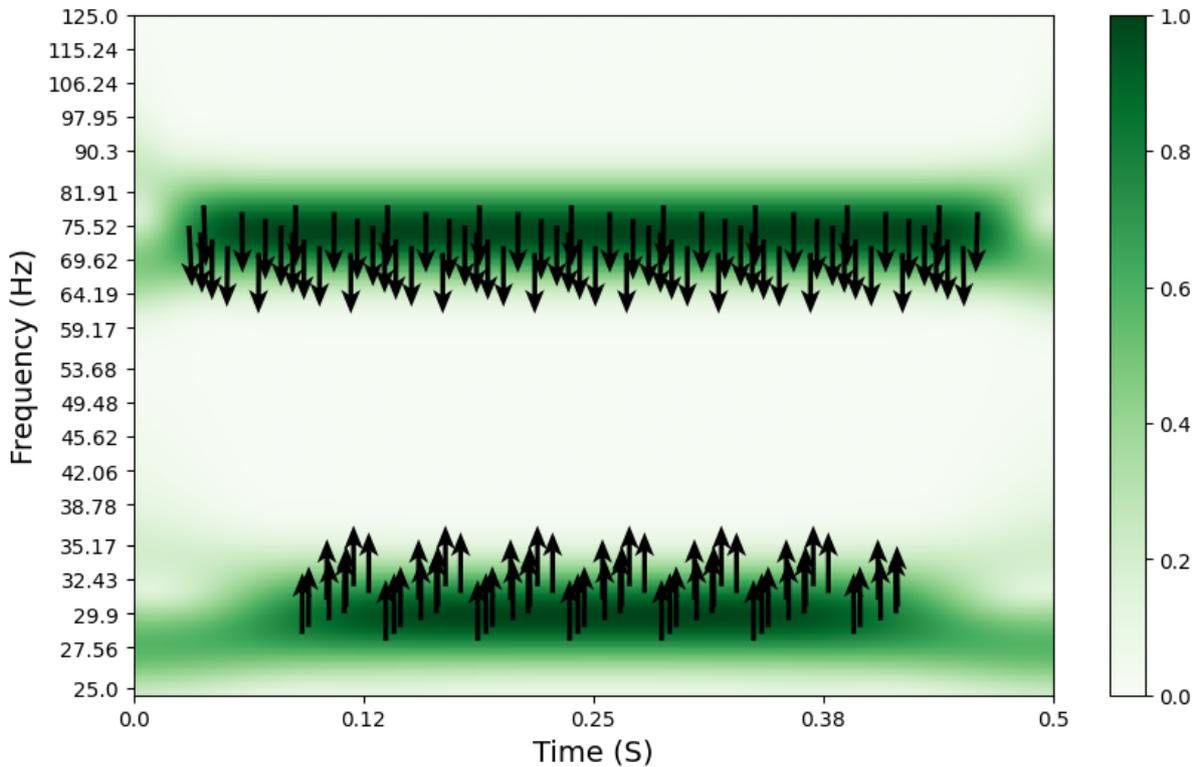

Figure B.2: Signals A and B Coherence.

To see a real world coherence example with the COI applied, examine Figure B.3. Here, we have the data of sea surface temperature (SST) anomalies in El Niño Region 3 (Niño 3), compared with the All-India Rainfall Index (AIRI), which is a typical proxy for monsoon performance in the Indian subcontinent. Both variables are sampled monthly from 1871 to 2003.

Figure B.4 shows the coherence between Niño 3 SST anomalies and AIRI (note that *period* lies on the y-axis in place of frequency, so the picture is upside-down when compared to previous figures). The plot reveals 2 bursts of wavelet coherence at periodicities between 2 and 7 years — the typical El Niño range. These coherent bands exhibit a consistent phase lag of about $3\pi/4$ to $\pi$ radians, indicating a delay of about 1 to 3.5 years between periods of sea warming in the equatorial Pacific, measured off the coast of South America, and monsoon rainfall over India. Despite the roughly 17,000 km separation, this suggests a large-scale correlation of oceanic temperature anomalies preceding atmospheric response by roughly 1 to 3.5 years.

The white dashed line overlaying Figure B.4 indicates the COI, underneath which the coherence coefficients are a valid representation of the co-movement between the SST anomalies and AIRI[1].

---

[1] The data and results in Figures B.3 and B.4 are sourced from the MATLAB Help Center's "Compare Time-Frequency Content in Signals with Wavelet Coherence" example [23].

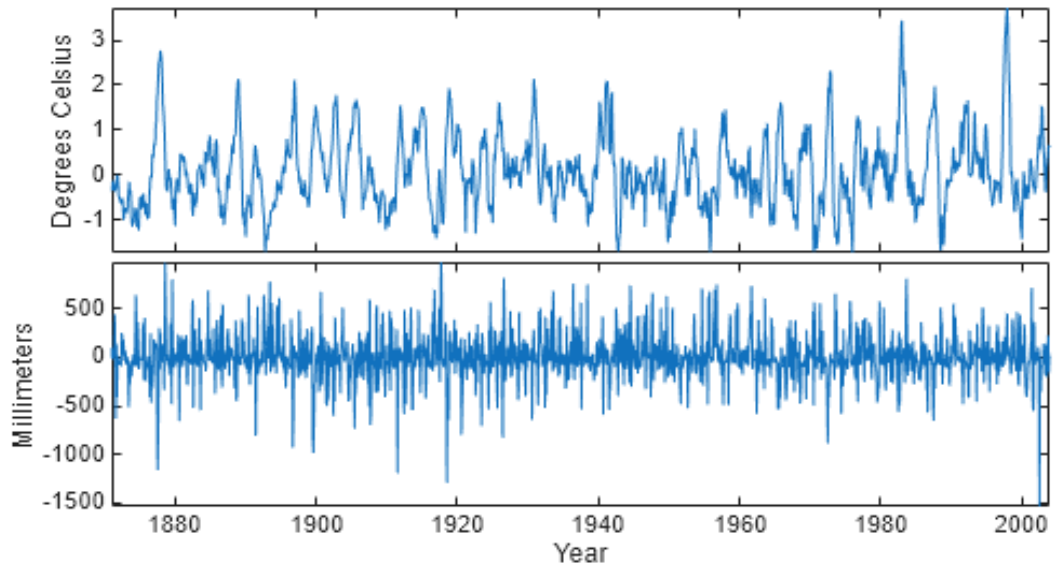

Figure B.3: Niño 3 – SST anomalies (Top) and deseasonalized AIRI (Bottom).

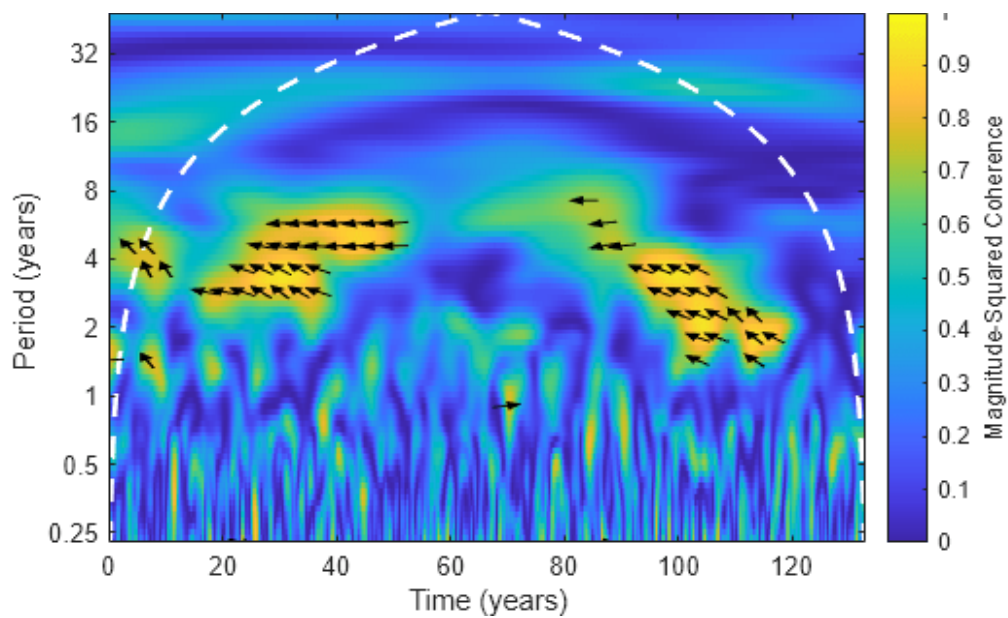

Figure B.4: Niño 3 and AIRI coherence array.

## B.2 Uncalibrated Coherence Results

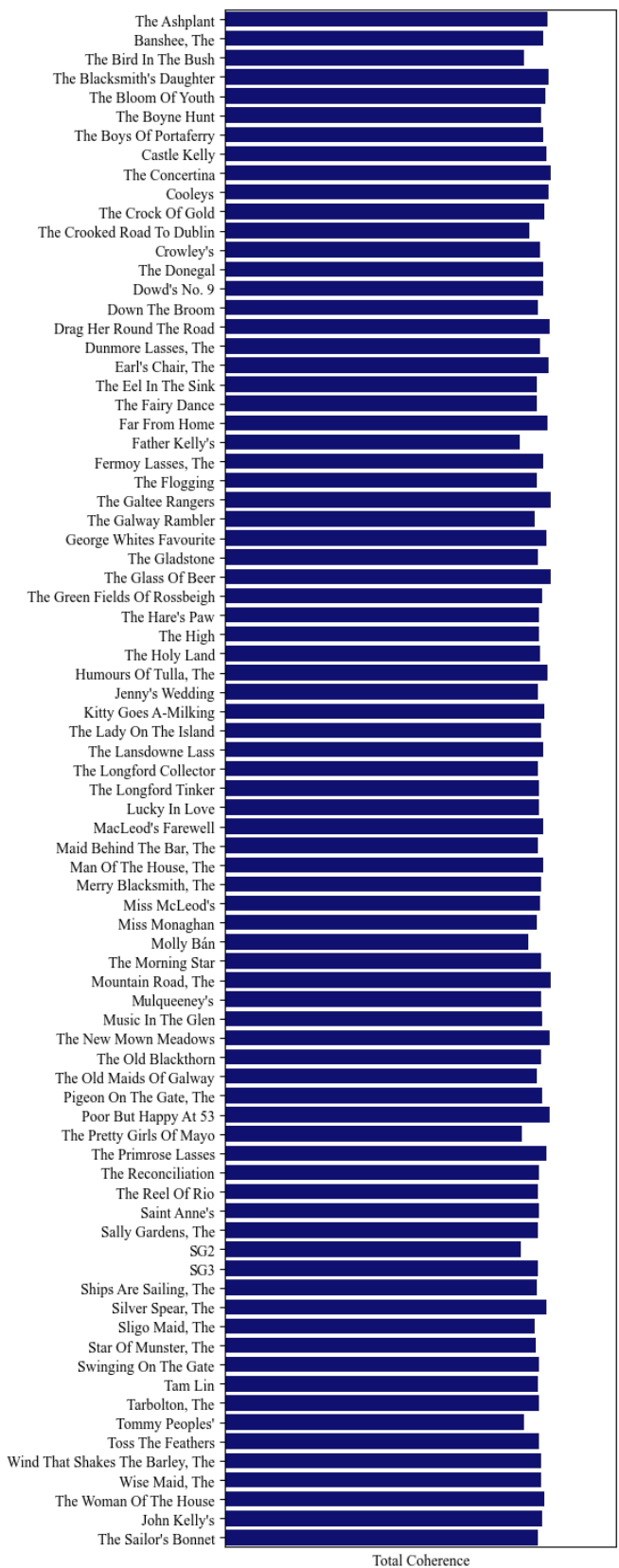

Figure B.5: Uncalibrated results of Control *a*.

For each coherence bar plot presented in this work (Figures 4.2, 4.4, and 4.10), the results have been calibrated relative to the tune with the lowest coherence value. This means that we have subtracted the coherence of the least coherent tune from all others, effectively zooming in on the relevant differences between the selected tune and the rest of the field.

One consequence of this calibration method is that it can make some tunes appear disproportionately advantaged when analyzing biases using control data. Figure B.5 displays the uncalibrated results from Control *a*, where we observe that all tunes yield roughly similar coherence values when paired with the blank signal.

While it is not so obvious which tunes carry an advantage for selection by the identification model, we can see that some tunes show quite little contribution to the coherence with the empty signal. These tunes include *The Bird in the Bush, Father Kelly's Reel, Molly Bán, The Pretty Girls of Mayo,* and *Tommy People's Reel.*